\documentclass[11pt, aps,english,nofootinbib, showkeys]{revtex4-1}
\usepackage[T1]{fontenc}
\usepackage[utf8]{inputenc}
\usepackage{xcolor}

\usepackage{babel}
\usepackage{pmboxdraw}
\usepackage{amstext}
\usepackage{graphicx}
\usepackage{bm}
\usepackage{amsmath}
\usepackage{makecell}
\usepackage{booktabs}
\usepackage[version=4]{mhchem}
\usepackage{float}
\usepackage{chemformula} 
\usepackage[T1]{fontenc} 
\usepackage{booktabs} 
\usepackage{makecell} 
\usepackage{xcolor}
\usepackage{comment}
\usepackage{booktabs}
\usepackage{array}
\usepackage{tabularx}
\usepackage{array} 
\usepackage{makecell} 

\begin{document}

\title{Unveiling the Design Rules for Tunable Emission in Graphene Quantum Dots: A High-Throughput TDDFT and Machine Learning Perspective}

\author{Mustafa Co\c{s}kun \"{O}zdemir}
\affiliation{Department of Chemistry, \.{I}zmir Institute of Technology,
\.{I}zmir, T\"{u}rkiye}

\author{Caner \"{U}nl\"{u}}
\affiliation{Department of Chemistry, \.{I}stanbul Technical University, \.{I}stanbul,
T\"{u}rkiye}

\author{\c{S}ener \"{O}z\"{o}nder}
\affiliation{Institute for Data Science \& Artificial Intelligence, Bo\u{g}azi\c{c}i University,
\.{I}stanbul, T\"{u}rkiye}
\email{Corresponding author: sener.ozonder@bogazici.edu.tr}

\begin{abstract}

The ability to tailor the optical properties of graphene quantum dots (GQDs) is critical for their application in optoelectronics, bioimaging and sensing. However, a comprehensive understanding of how shape, size and doping influence their emission properties remains elusive. In this study, we conduct a systematic high-throughput time-dependent density functional theory (TDDFT) and machine learning analysis of 284 distinct GQDs, varying in shape (square, hexagonal, amorphous), size ($\sim$1–2 nm) and doping configurations with elements B, N, O, S and P at varying concentrations (1.5–7\%). Our findings reveal clear design principles for tuning emission wavelengths based on dopant type, concentration and GQD geometry. Notably, sulfur doping at specific concentrations consistently results in higher emission energies, with certain configurations yielding emissions within the visible range. By elucidating how quantum confinement effects, symmetry breaking and dopant-induced modifications govern GQD optical properties, we provide practical design rules for tailoring emission spectra for next-generation optoelectronic, bioimaging and sensing applications.
\end{abstract}

\maketitle

\section{Introduction}

Graphene quantum dots (GQDs) are carbon-based 2D nanomaterials which consist of sp$^2$ hybridized carbon atoms with lateral dimensions typically less than 10 nm, displaying unique electronic, optical and chemical properties due to quantum confinement and edge effects \cite{Mahto2024, Reckmeier2016, Kovalchuk2015, Deng2014, Du2023, Mandal2012}. These carbon-based 2D nanomaterials exhibit exceptional photoluminescence, high surface area and biocompatibility, making them ideal candidates for various applications in bioimaging, sensing and optoelectronics \cite{Tian2018, Sun2013, Yuan2018, Wang2021, Singh2020, Zou2016, Omer2019, Saikia2022, Li2022, Lim2015, Lee2021, Cui2022}. 

GQDs have become center of attention in recent years due to their tunable fluorescence properties, which can be controlled by altering their size, shape, composition and functional groups, facilitating their use in advanced light-emitting devices, solar cells and photodetectors \cite{Tian2018, Sun2013}. Two primary pathways contribute to the fluorescence of GQDs: intrinsic state emission, which is associated with the sp$^2$-hybridized carbon core and surface state emission, which is influenced by the surface chemical groups and carbon bonds \cite{Xu2020}. The combination of surface defects, functional groups and heteroatom doping in the core of GQDs greatly affects their fluorescence properties \cite{Xu2020}. Additionally, surface alterations like oxidation and reduction can adjust the colors of their emissions \cite{Xu2020, Wang2018, Yang2020}.

Experimental research on heteroatom doped graphene quantum dots has shown that doping distinct atoms to GDQs can control a variety of optical properties of the dots, including photoluminescence wavelength and quantum yield. The photophysical properties of GQDs undergo significant changes when doped with N and B atoms such as increasing their quantum yield and resulting in controllable shifts in their emission wavelength \cite{Liu2017, Budak2020, Li2014}. It was demonstrated that the photoluminescence of GQDs can be enhanced by doubly doping them with Mg and N \cite{Li2014}.   It was also shown that doping GQDs with S, N and B resulted in red light emission \cite{Liu2017}. In a separate recent study, doping GQDs with N and P or B can create dual emission spectrum \cite{Parvin2017, Budak2021}. Enhanced photoluminescence, improved photostability and tailored electronic properties can be obtained with heteroatom-doped GQDs. These can in turn result in important breakthroughs in bioimaging, energy storage, optoelectronics, environmental monitoring and other fields \cite{Miao2020, Fernandez2023, Xu2016, Sohal2021}.

Time-dependent density functional theory (TDDFT) has been successfully applied to calculate excitation energies, optical absorption and emission spectra and electronic transition properties of molecular systems. It offers a compelling alternative to computationally demanding, highly correlated ab initio methods for investigating the electronic structure and optical properties of molecular systems \cite{Runge1984, Sarkar2021, Wang2022}. By extending the principles of ground-state density functional theory (DFT) to the time domain, TDDFT achieves this while maintaining acceptable accuracy at a lower computational cost. The versatility and efficiency of TDDFT have made it a particularly attractive approach for studying the unique characteristics of GQDs. Numerous studies have utilized TDDFT to investigate the optical properties of GQDs, exploring topics such as the impact of nitrogen dopant configuration \cite{Yang2020}, the utilization of enhanced light absorption \cite{Wazzan2022}, the influence of edge modifications\cite{Jabed2021} and more \cite{Wang2018, Ozonder2023, Schumacher2011, Zhao2014, Liu2024}.

The TDDFT literature usually consists of studies concerning few particular types of GQDs and these fragmented studies do not reveal the systematics of the effect of shape, size and dopants on the emission spectrum of GQDs. In this work, we conduct a high-throughput TDDFT calculations aiming to investigate the effect of shape, size, (co-)dopant type and dopant percentage on the emission spectrum of 284 distinct GQDs. The systematics and trends to be learned from this analysis can be used as a guide for design engineering desired emission spectra for various applications and concurrently  reducing the number of trial-and-errors during synthesis.

\section{Computational Details}

\begin{figure*}[!t]
\includegraphics[scale=0.7]{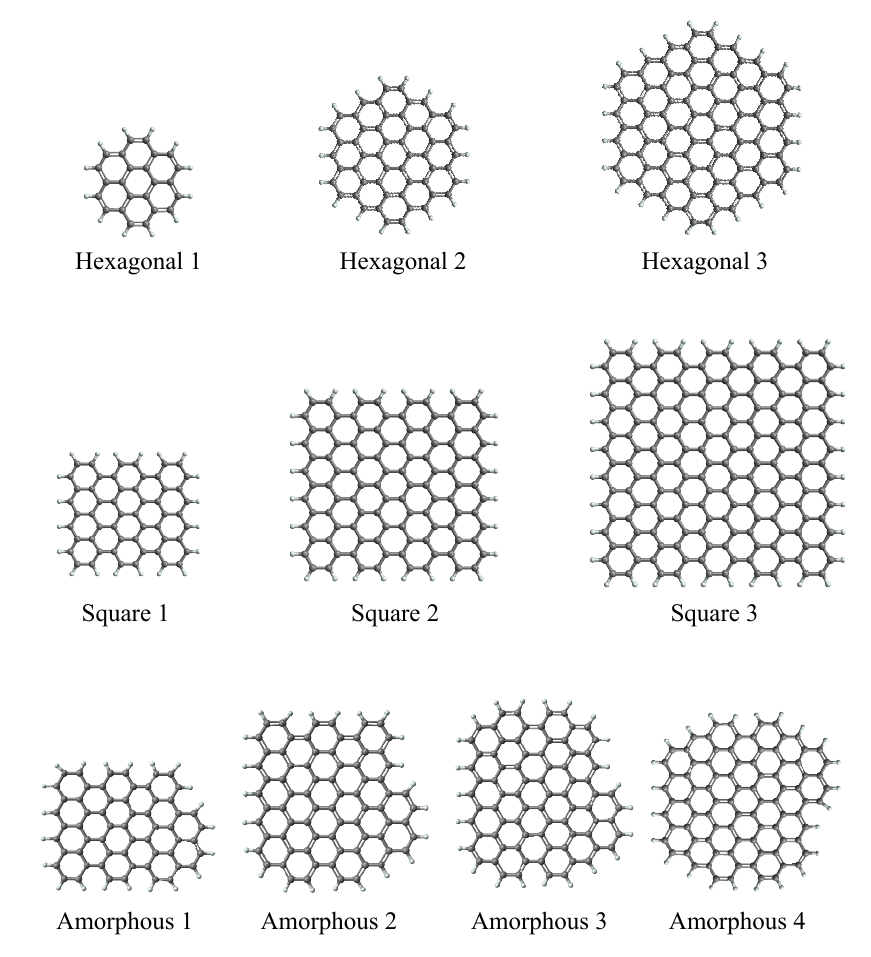}

\caption{The structures along with their singly- and doubly-doped variants were used to calculate emission energies via TDDFT. The square shapes have 1 nm, 1.5 nm and 2 nm side lengths.}

\label{structures}
\end{figure*}

Fluorescence emission energies from S1$\rightarrow$S0 for square, hexagonal and amorphous-shaped GQDs have been calculated. The ten structures considered are shown in Figure \ref{structures}. These pristine GQDs and their singly-doped (B, O, P, S, N) and doubly-doped variants (BO, OS, BP, BS, NO, NS, BN, NP, OP, SP) at dopant percentages of 1.5\%, 3\%, 5\% and 7\% constitute a total of 610 GQDs. This large sample size extensively covers the chemical space of GQDs with sizes of approximately 1-2 nm.

Amorphous-shaped GQDs structurally interpolating hexagonal and square shapes are derived from 1 nm square structures to examine the effect of structural defects on the emission spectrum. Furthermore, the amorphous structures facilitate the investigation of the effects on the emission energies caused by the breaking of  symmetries present in hexagonal and square structures. Symmetric structures contain degenerate states that affect the HOMO-LUMO gap, hence, symmetry breaking is of immense effect on the emission energies.

Hydrogen atoms are added to the edges of the molecules to passivate the dangling bonds. All structures are ensured to be neutral in charge and have zero total spin; an extra dopant or hydrogen atom is added to or removed from some structures to achieve this. The reported dopant percentages are defined as the percentage of dopant atoms relative to all atoms excluding hydrogens.

Geometries have been pre-optimized using the Semiempirical Extended Tight-Binding Program Package \cite{Bannwarth2020} at the Bannwarth2019-xTB \cite{Bannwarth2019} level of theory to reduce the initial computational cost. Ground state (S0) and excited state (S1) optimizations have been performed with Gaussian16 \cite{g16} using the hybrid functional B3LYP \cite{Becke1993,Lee1988,Vosko1980} and the basis set 6-31G(d) \cite{Dill1975,Ditchfield1971,Francl1982,Frisch2009,Gordon1982,Hariharan1973,Hehre1972}. Both DFT for geometry optimization and TDDFT for excited state calculations are performed in water within the framework of the polarizable continuum model (PCM) \cite{Improta2007,Becke1993}. Both of the optimized S0 and S1 states have been verified by frequency analysis to be free from imaginary frequencies.

Structures that did not converge, were not optimized or had virtual frequencies have not been re-run and consequently not reported, as they would require manual intervention, which in turn impedes the streamlined workflow of high-throughput screening. This yielded 284 successful TDDFT results out of 610 possible structures. All calculations were done with 128-core nodes and the total wall time for the entire calculations reported here is 6223 node-hours.

\begin{figure*}[!h]
\includegraphics[width=0.92\textwidth]{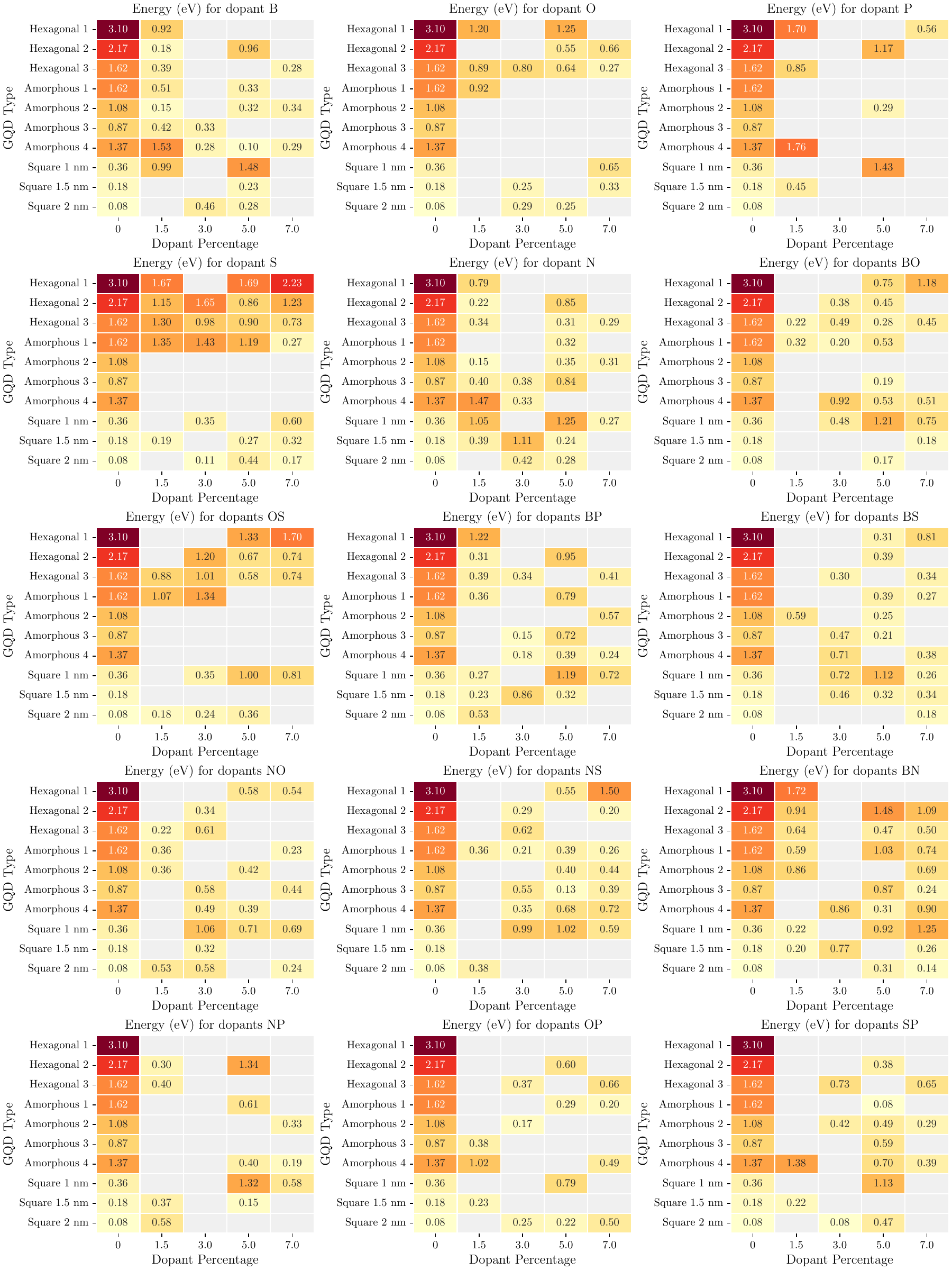}

\caption{Emission energies in eV of 284 GDQs are presented where each heatmap panel for a different single or double dopants. Missing slots belong to the structures that did not converge, were not optimized or had virtual frequencies.}

\label{dopantsgrid}
\end{figure*}

\section{Results and Discussion}

The electronic structure of GQDs and accordingly their emission energies and intensities (oscillator strength) vary strongly depending on the nanocrystal's shape, size, dopant type and dopant percentage. In this work, the emission energies of 284 different structures were found using TDDFT excited-state calculations. All emission energies in eV are summarized in Figure \ref{dopantsgrid}, where each heatmap panel belongs to a different dopant atom(s) (The molecular structures, emission energies and their corresponding oscillator strengths are provided in the Supplementary Material.). According to the Kasha's rule, fluorescence emission can only occur from the lowest energy state of S1 state. The energies reported in Figure \ref{dopantsgrid} corresponds to S1$\rightarrow$S0 transitions. Possible phosphorescence phenomenon which involves triplet states have not been investigated in this study.  

As revealed in Figure \ref{dopantsgrid}, for the pristine hexagonal and square structures, the emission energy decreases with increasing structure size as expected due to  quantum confinement. As the structure size grows, its HOMO-LUMO gap closes, causing the photoluminescence spectrum to shift towards the IR region. This trend is somewhat preserved for the doped hexagonal and square structures. Amorphous structures were specifically prepared to be a transitional form between square and hexagonal ones to check if their energies interpolate between those of square and hexagonal structures. This indeed can be seen from Figure \ref{dopantsgrid}.

Hexagonal 2 (\ch{C54H18}) and Square 1 (\ch{C54H20}) have the same number of carbon atoms. Hexagonal 3 (\ch{C96H24}) and Square 2 (\ch{C104H28}) also have a similar number of atoms. However, these structure have stark differences in terms of emission energy. This can be attributed to the fact that they have different symmetry properties. The pristine square structure has $D_{4h}$ point group symmetry, while the pristine hexagonal structure has $D_{6h}$ point group symmetry. The amount of degeneracy in orbital energies affects the HOMO and LUMO energies, as well as the gap between them. This can be related to the particle in a 2D square and circular box toy models from introductory quantum mechanics. In a 2D box, energy depends on the quantum numbers as $E \propto n_x^2 + n_y^2$, where $n_x=n_y$ in the square box case, which in turn brings degeneracy. In a circular box, quantum numbers are $x_{mn}$, which is the $n$-th zero of the $m$-th Bessel function $J_m$. So, despite having the same or similar atomic content, different symmetries of shapes yield different emission energies.

\begin{table}[!h]
\centering
\small 
\caption{Structures with emission energy in the visible region. Dopant colors are P (orange), B(pink), N (blue), O (red) and S (yellow).}
\label{visibletable}
\begin{tabular}{@{} p{0.5\textwidth} @{} p{0.5\textwidth} @{}}
\begin{tabular}{@{} >{\centering\arraybackslash}m{1.7cm} >{\centering\arraybackslash}m{1.5cm} >{\centering\arraybackslash}m{1.5cm} >{\centering\arraybackslash}m{2.5cm} @{}}
\toprule
Structure & Energy (eV) & Wavelength (nm) & Dopants (\%) \\
\midrule
\includegraphics[height=1.7cm]{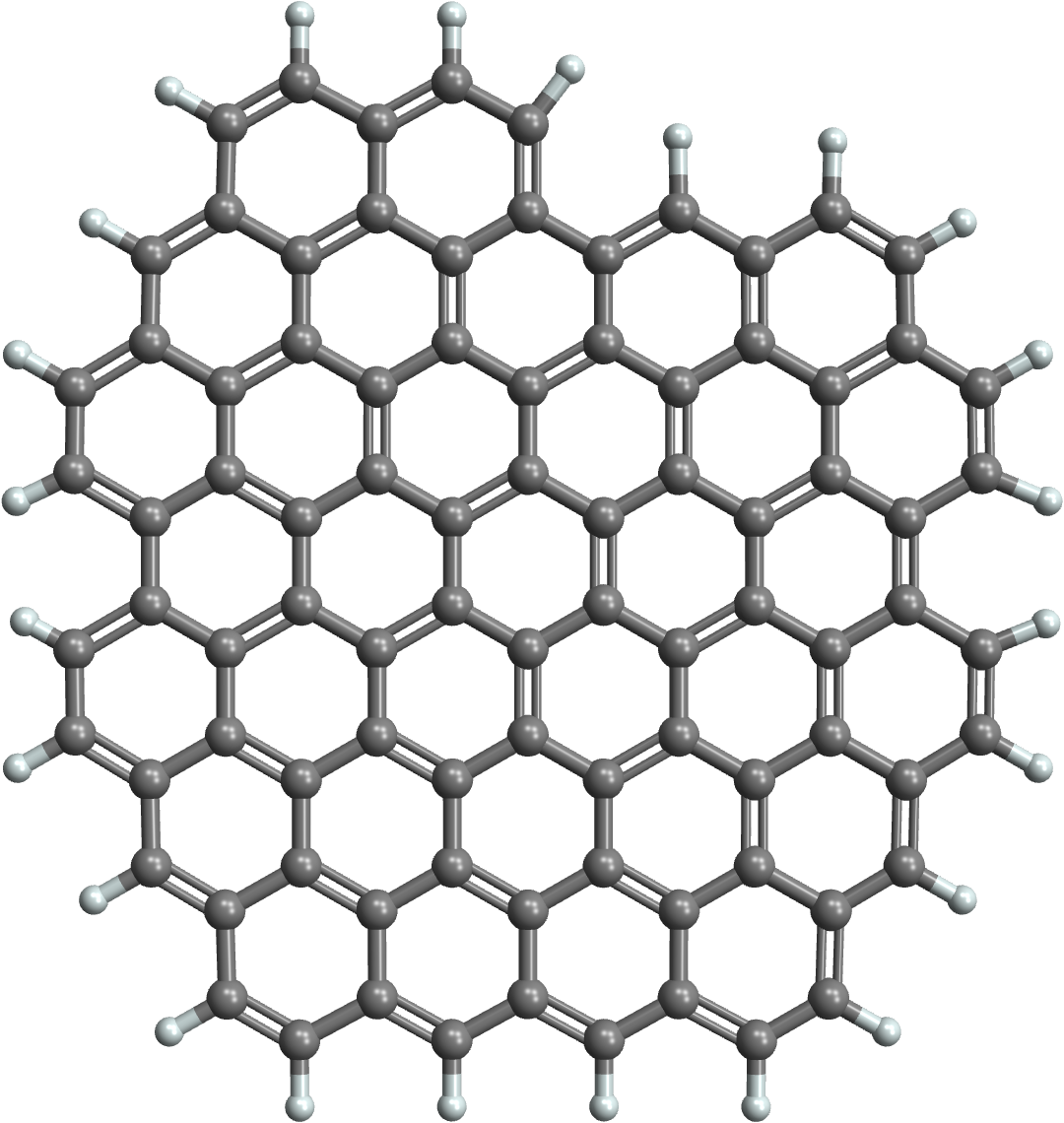} & 1.62 & 763 & pristine \\
\includegraphics[height=1.7cm]{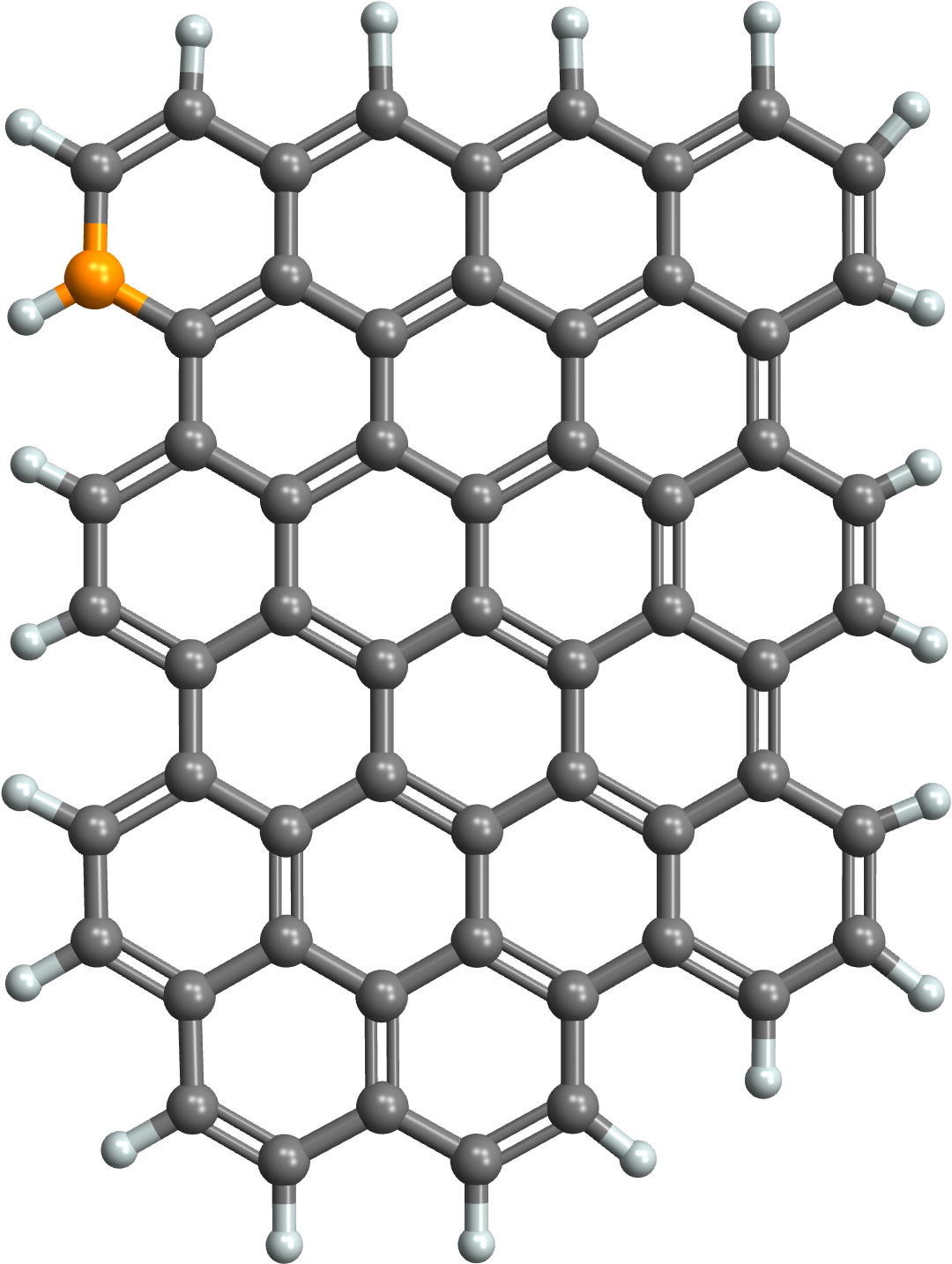} & 1.76 & 705 & P (1.5\%) \\
\includegraphics[height=1.7cm]{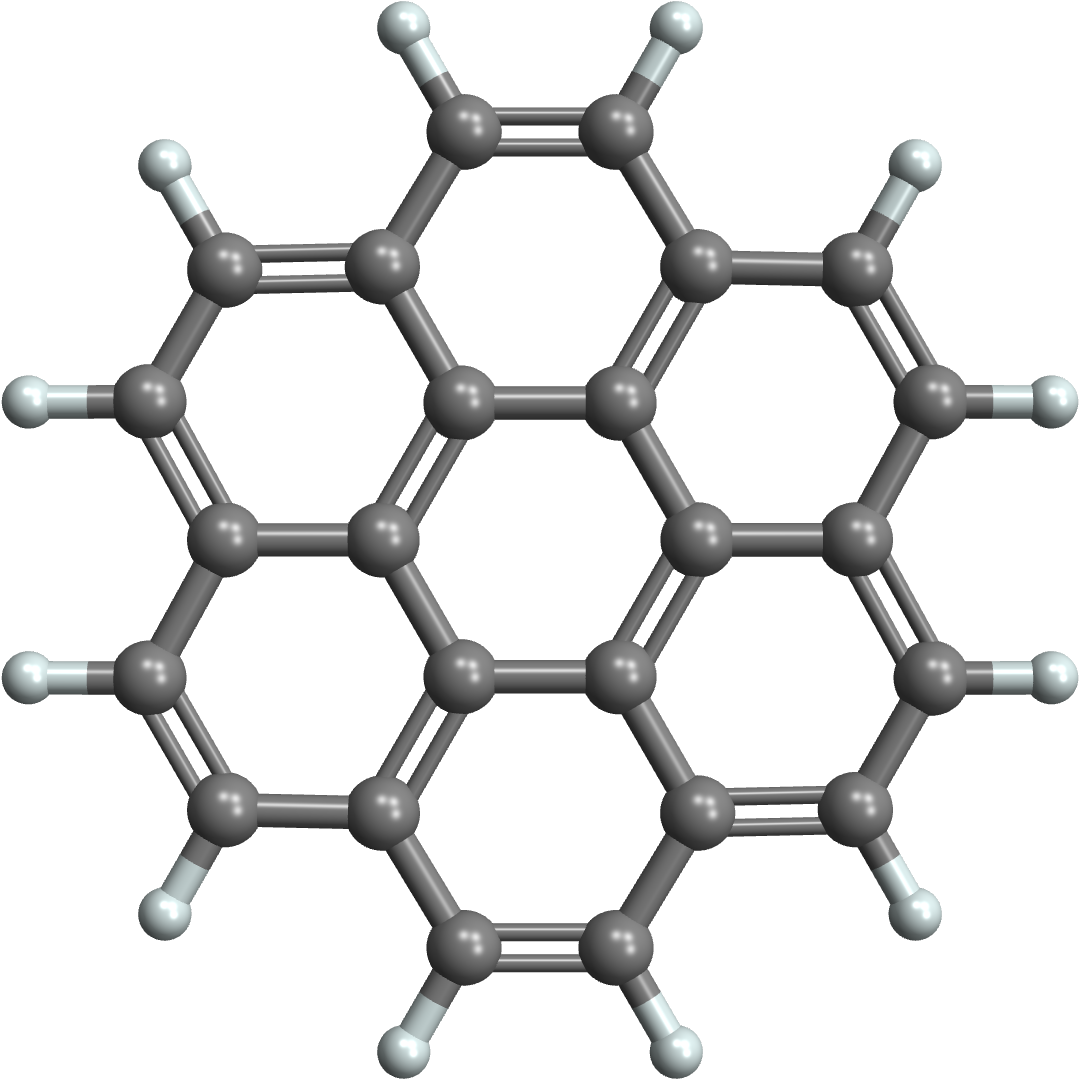} & 3.10 & 399 & pristine \\
\includegraphics[height=1.7cm]{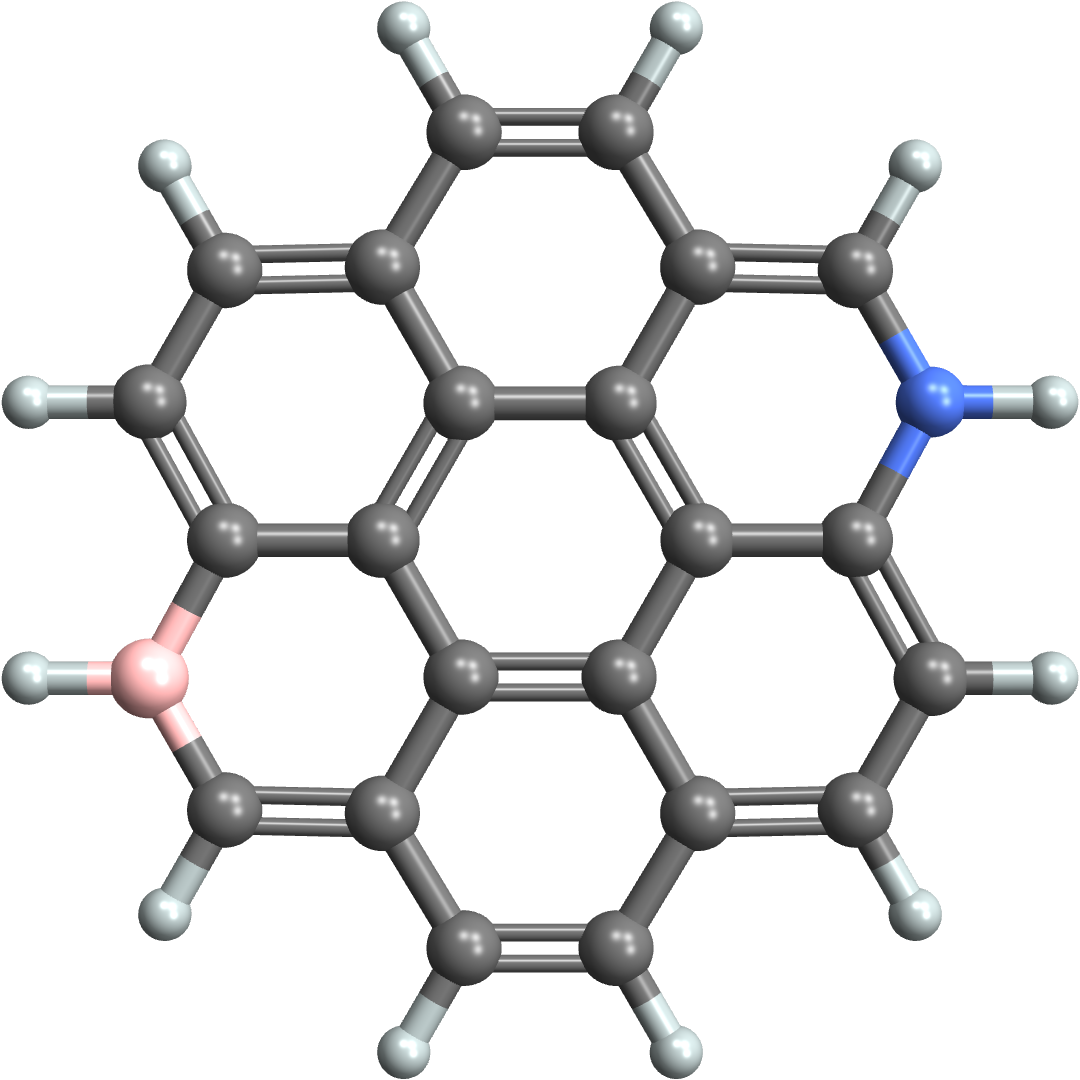} & 1.72 & 722 & BN (1.5\%) \\
\includegraphics[height=1.7cm]{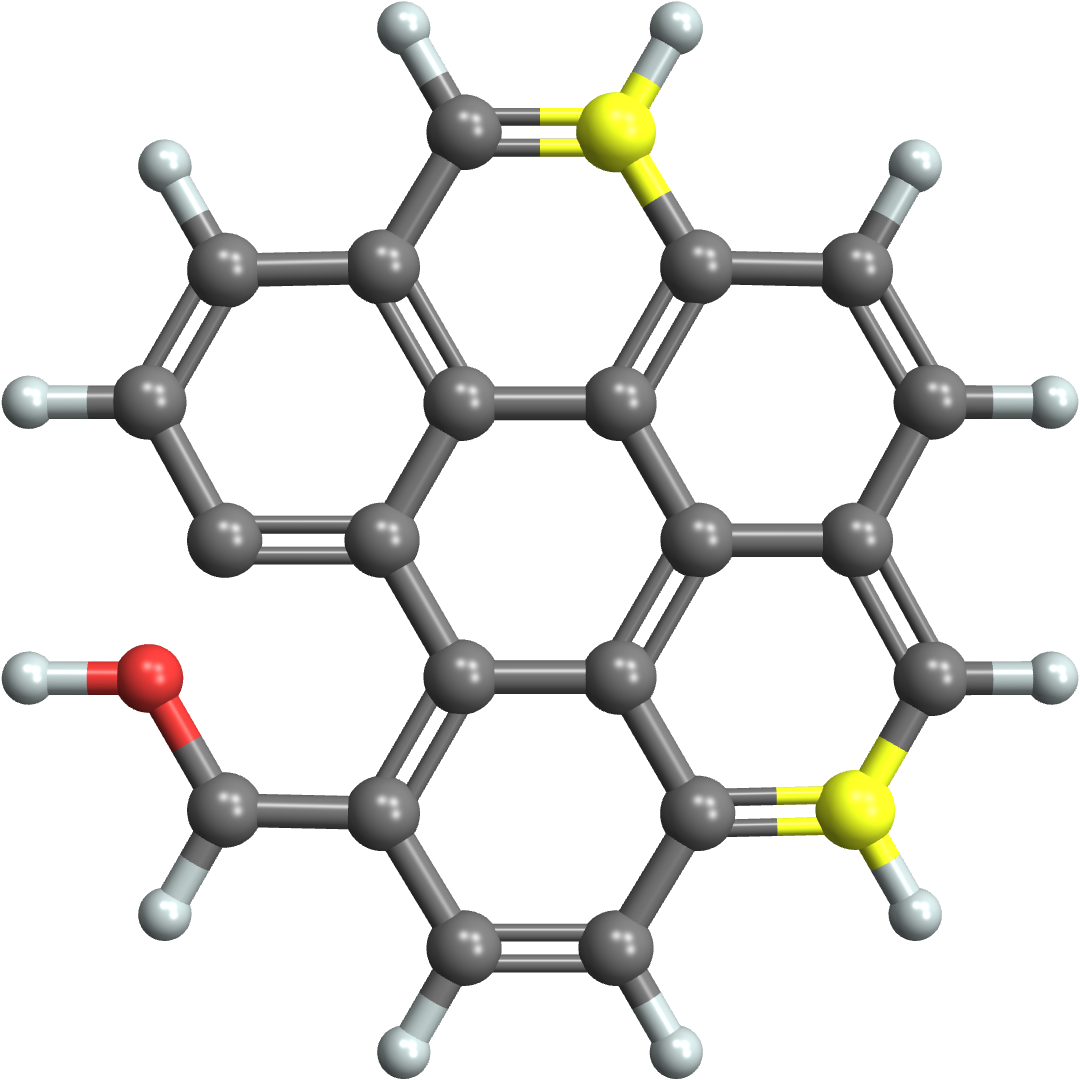} & 1.70 & 728 & OS (7\%) \\
\includegraphics[height=1.7cm]{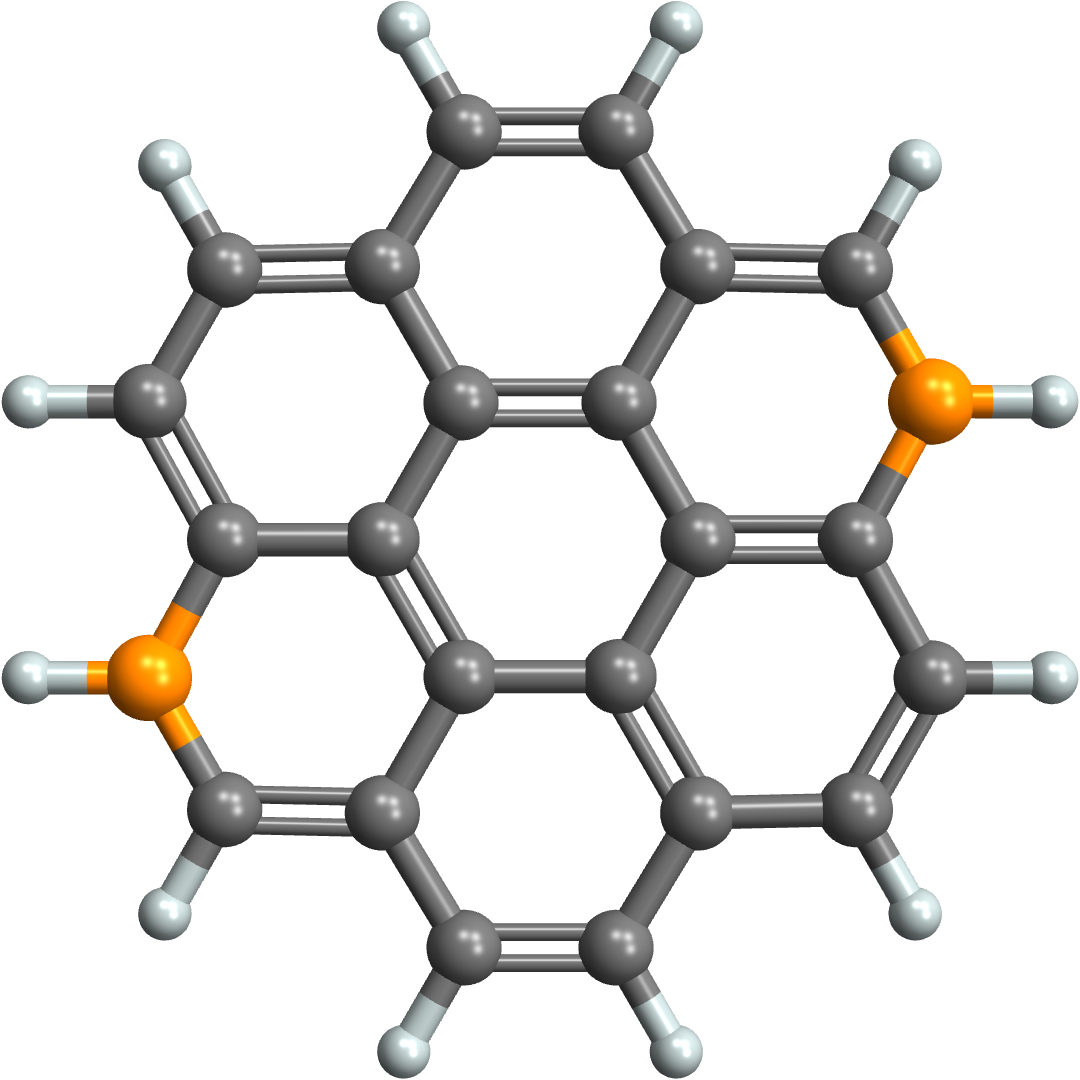} & 1.70 & 728 & P (1.5\%) \\
\bottomrule
\end{tabular}
&
\begin{tabular}{@{} >{\centering\arraybackslash}m{1.7cm} >{\centering\arraybackslash}m{1.5cm} >{\centering\arraybackslash}m{1.5cm} >{\centering\arraybackslash}m{2.5cm} @{}}
\toprule
Structure & Energy (eV) & Wavelength (nm) & Dopants (\%) \\
\midrule
\includegraphics[height=1.7cm]{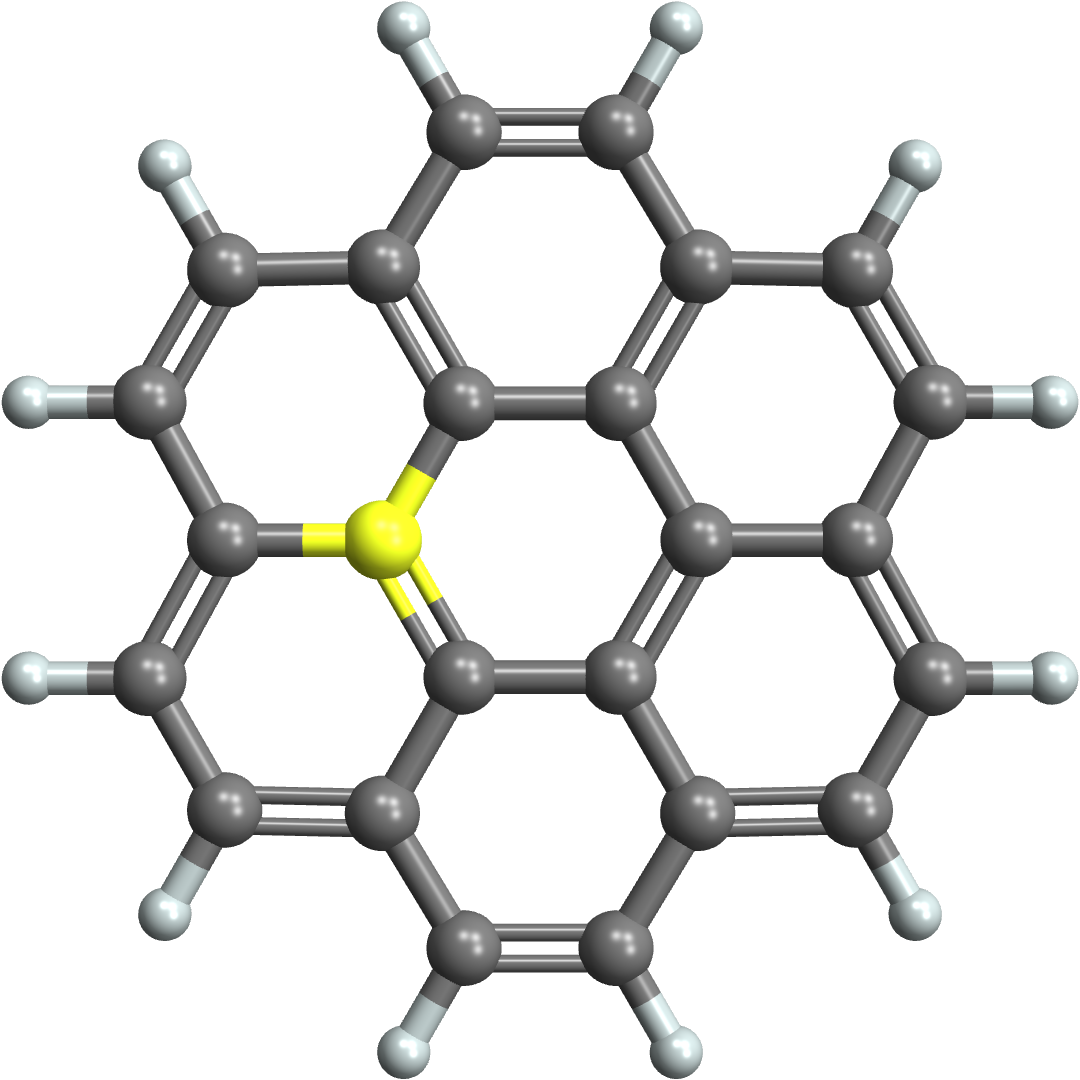} & 1.67 & 740 & S (1.5\%) \\
\includegraphics[height=1.7cm]{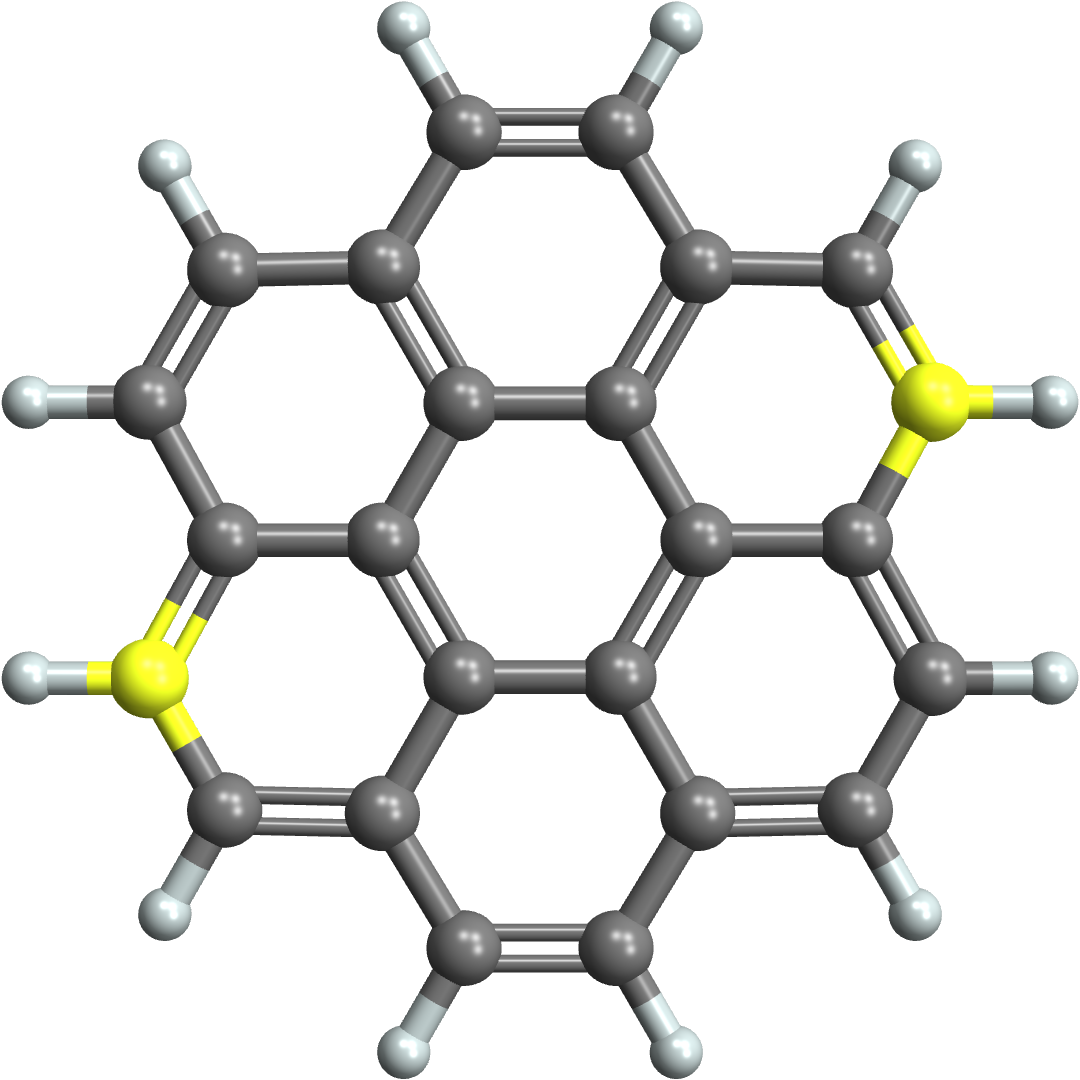} & 1.69 & 734 & S (5\%) \\
\includegraphics[height=1.7cm]{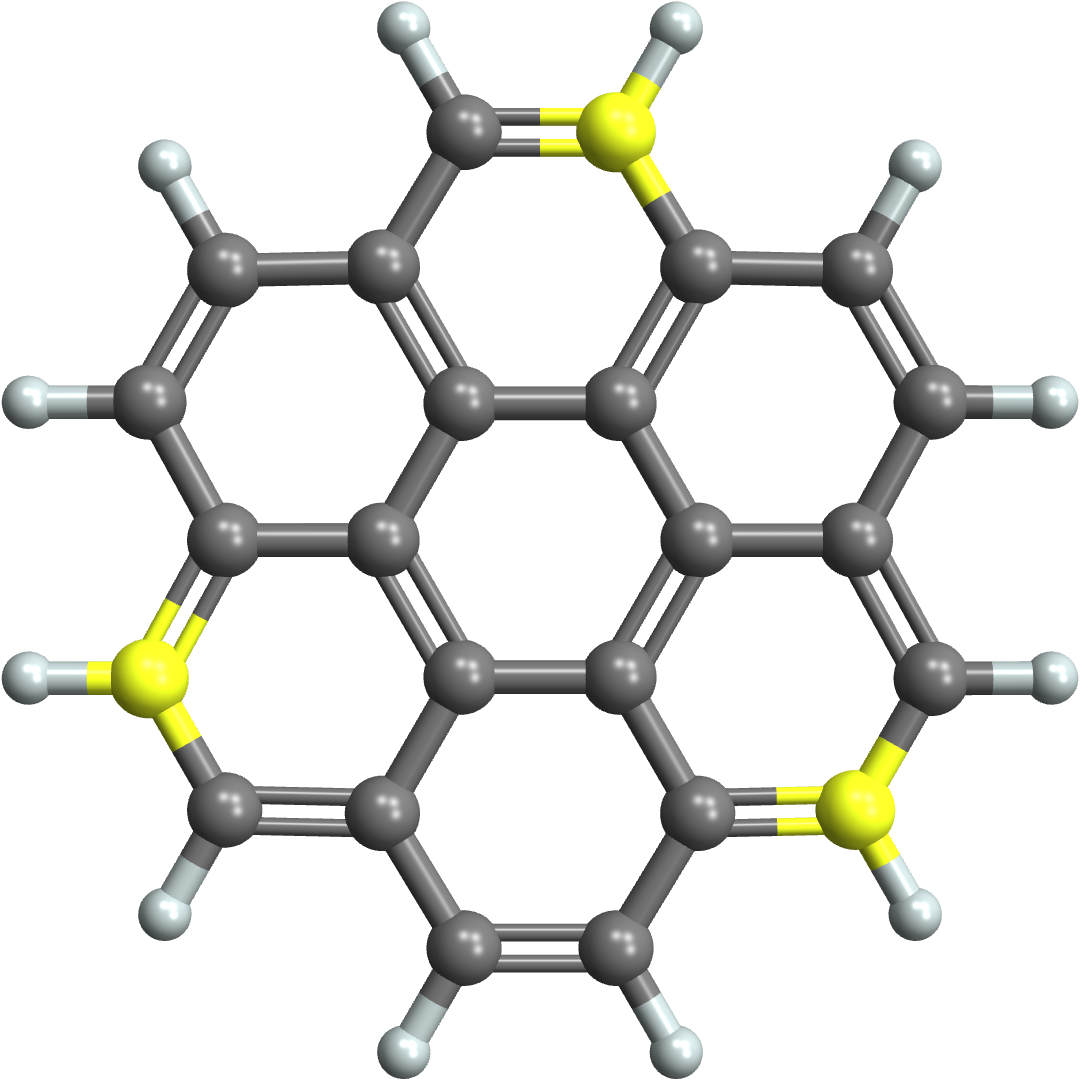} & 2.23 & 556 & S (7\%) \\
\includegraphics[height=1.7cm]{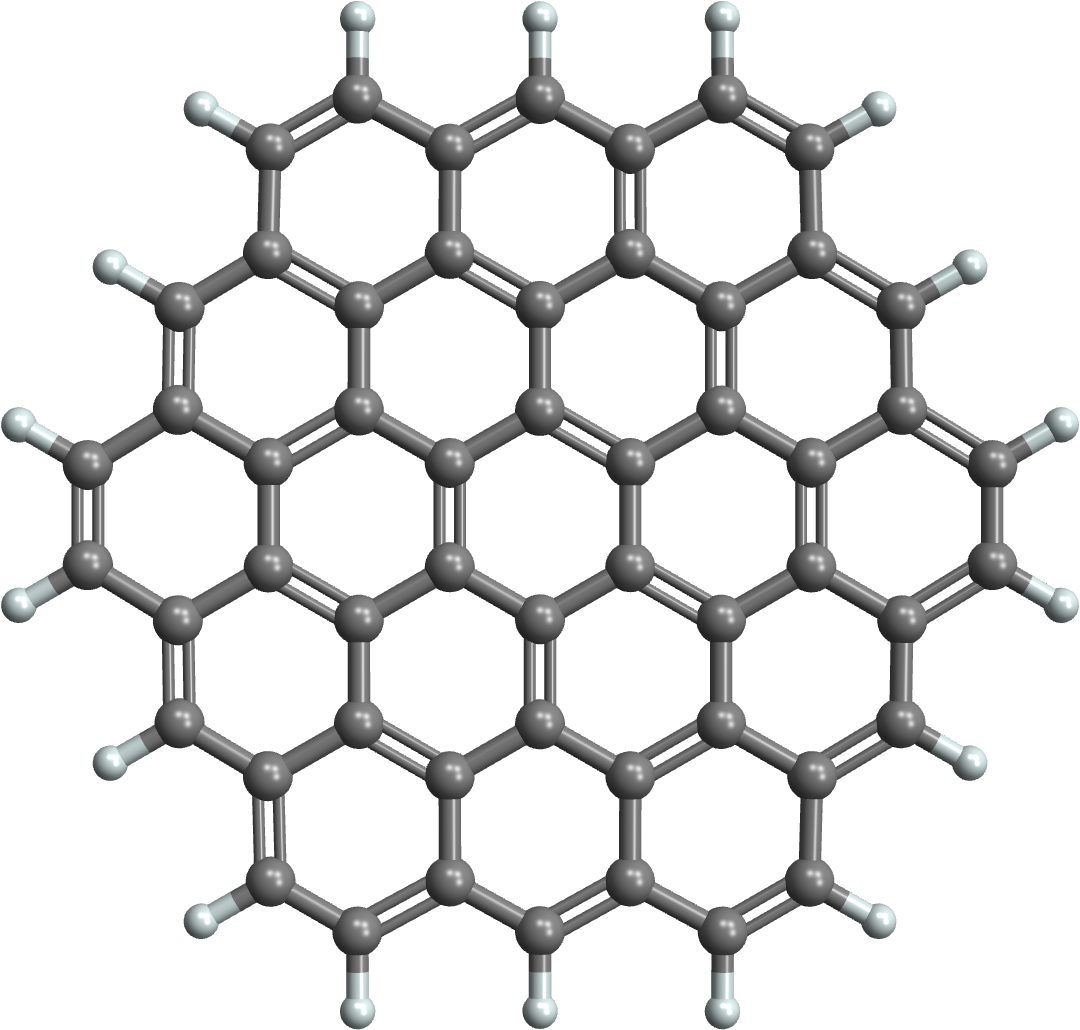} & 2.17 & 572 & pristine \\
\includegraphics[height=1.7cm]{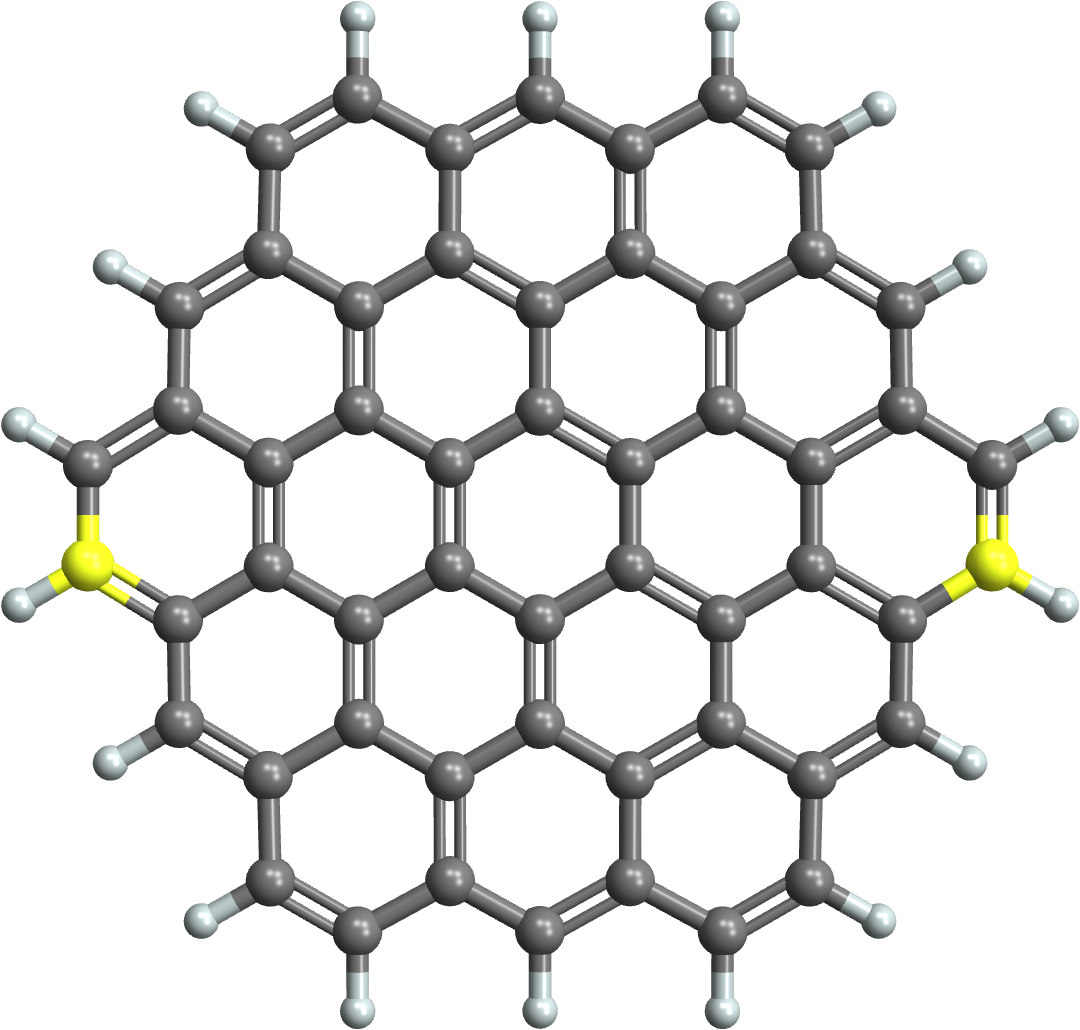} & 1.65 & 750 & S (3\%) \\
\includegraphics[height=1.7cm]{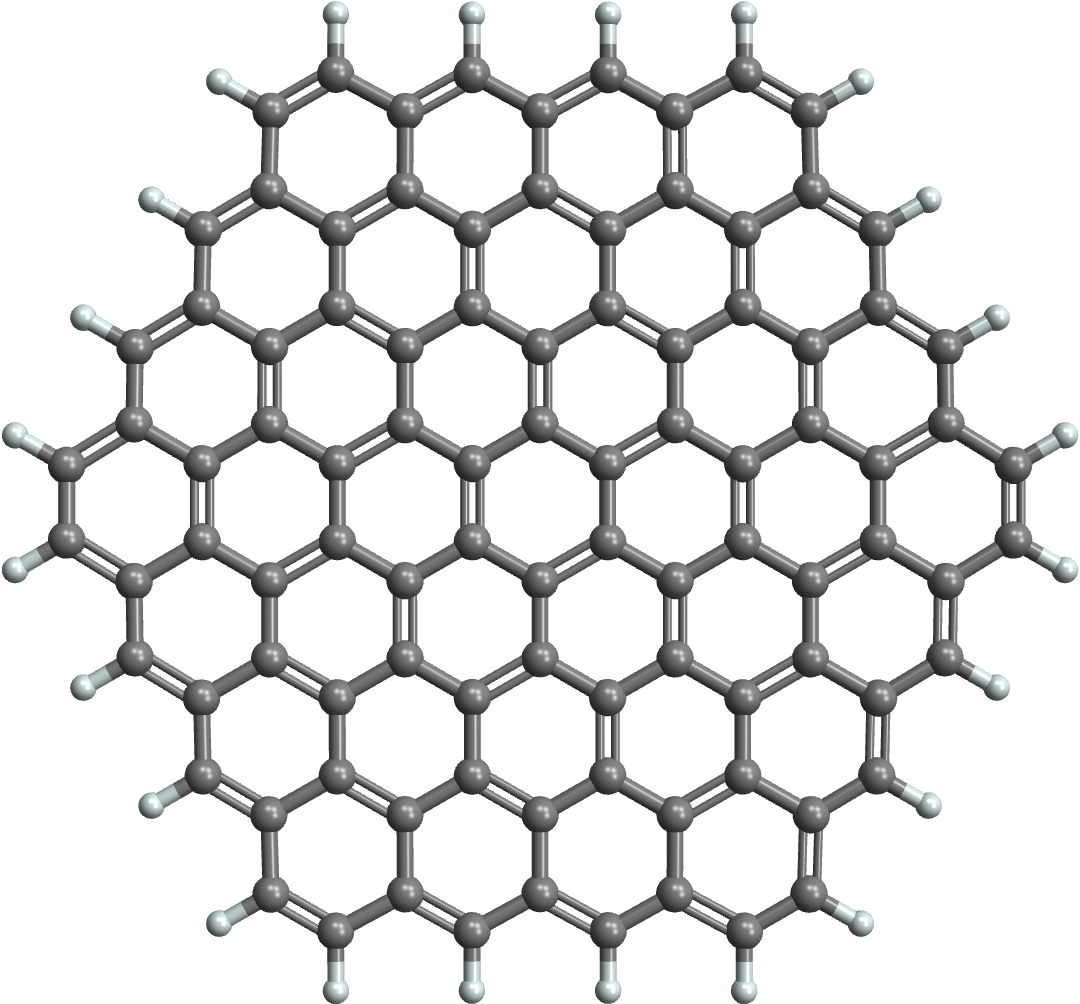} & 1.62 & 767 & pristine \\
\bottomrule
\end{tabular}
\end{tabular}
\end{table}

When we specifically examine the direct effects of dopants on hexagonal and amorphous structures (Figure \ref{dopantsgrid}), we find that sulfur as the dopant usually leads to the highest emission energies. Sulfur, by providing extra valence electrons in the $3s^2 3p^4$ orbitals, facilitates N-type doping on the carbon surface. Although sulfur and oxygen have the same number of valence electrons, the valence electrons of oxygen are located in the second shell and do not result in high emission energies unlike in the case of sulfur. Additionally, doping with oxygen or codoping with oxygen and sulfur atoms yields similar results. We also observe that sulfur-doped structures predominantly have emission energies above 1 eV. The calculated emission energies range between 0.08 and 3.10 eV. Table \ref{visibletable} shows the only 12 structures that emit light in the visible electromagnetic spectrum whereas the majority of GQD structures emit in the infrared (IR) region.

\begin{figure*}[h]
\includegraphics[scale=0.3]{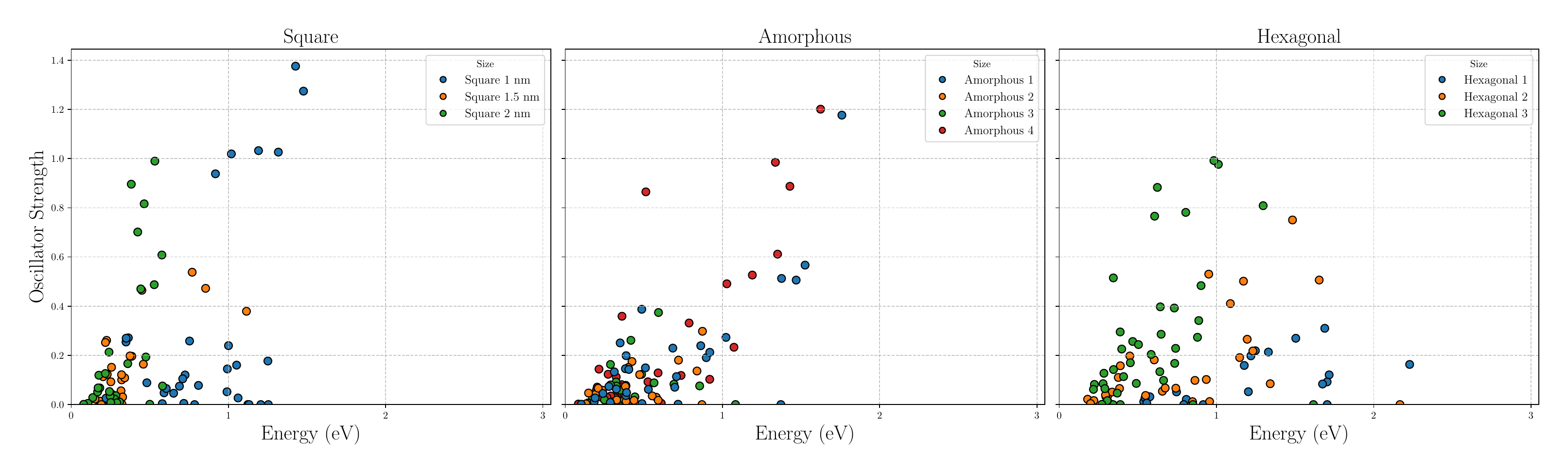}

\caption{Oscillator strength of 284 GQDs with respect to their shape and emission energies as calculated from TDDFT.}

\label{sizes}
\end{figure*}

Figure \ref{sizes} shows the energy and oscillator strength of the 284 structures, categorized into groups of square, amorphous and hexagonal. Most emission energies are in the infrared range or beyond, while some fall within the visible range (1.6-3.3 eV or 380-750 nm). A few hexagonal structures are found to have emission energies exceeding 2 eV.

We now transition into the machine learning analysis of the GQD dataset to extract new insights. Hierarchical clustering (HC) is employed here to group GQDs based on their properties and emission energies. In this study, we applied Ward's method to compute the distances between GQDs, as it minimizes the variance within clusters. 

\begin{figure*}[t]
\includegraphics[scale=0.5]{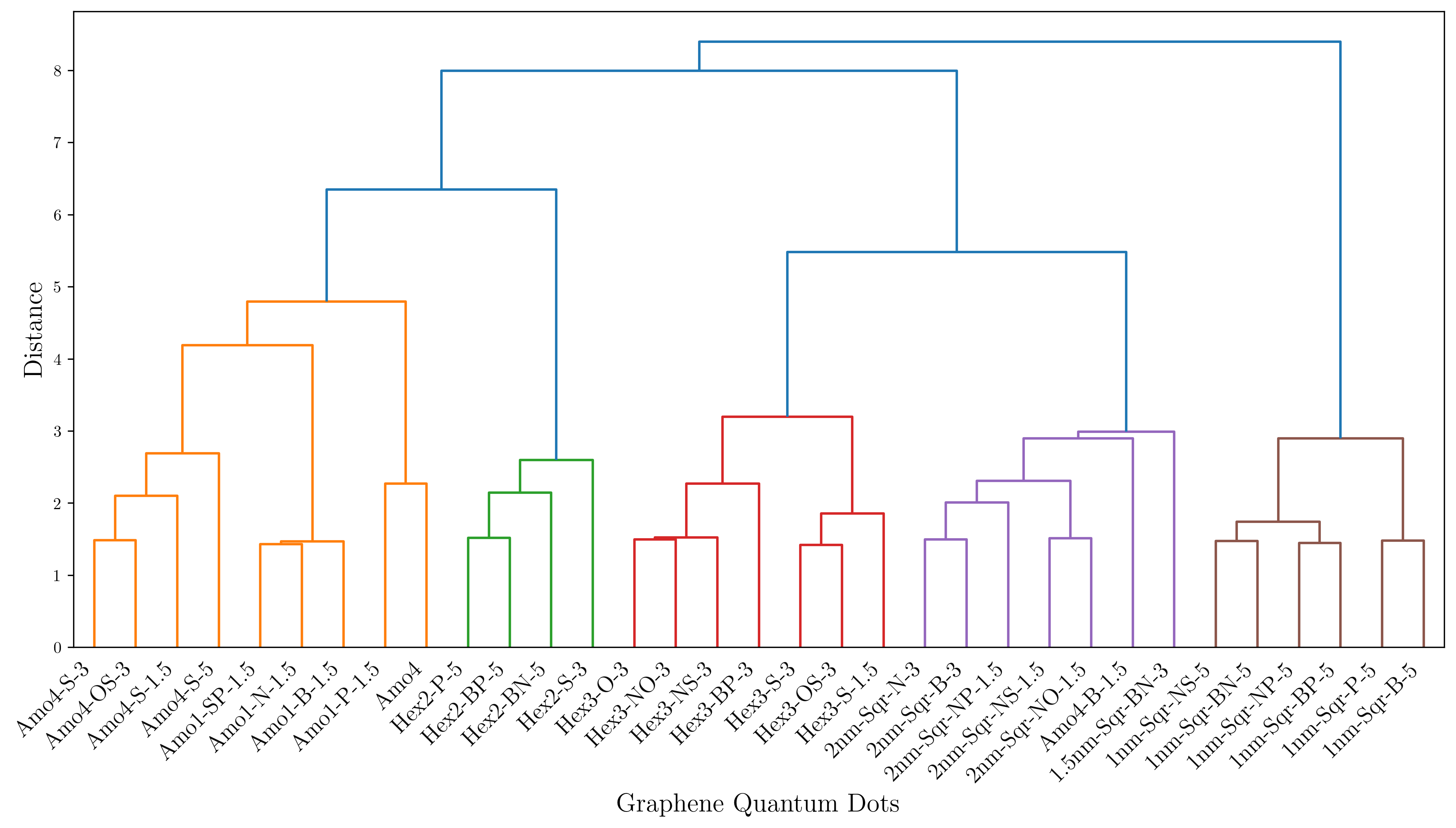}

\caption{Dendrogram plot of GQD data from hierarchical clustering modeling. Only the structures that have oscillator strength greater than 0.5 are kept. The last numbers in the names of the structures correspond to dopant percentages.}

\label{hc}
\end{figure*}

The hierarchical clustering dendrogram in Figure \ref{hc} reveals significant relationships between the structural configurations and dopant types. For example, shape and size appear to be the most important factors in determining similarity. However, some Hexagonal 2 structures cluster with certain Amorphous 1 and 4 structures rather than with Hexagonal 3 structures. This suggests that dopant type and percentage may outweigh the effects of shape and size. The emergence of mixed clusters involving both amorphous and hexagonal shapes may be attributed to the complex interplay of symmetry-breaking effects, dopant types, and quantum confinement. These clusters also demonstrate how specific dopants, such as sulfur, nitrogen, and phosphorus, significantly influence the clustering of GQDs.

\begin{figure*}[t]
\includegraphics[scale=0.5]{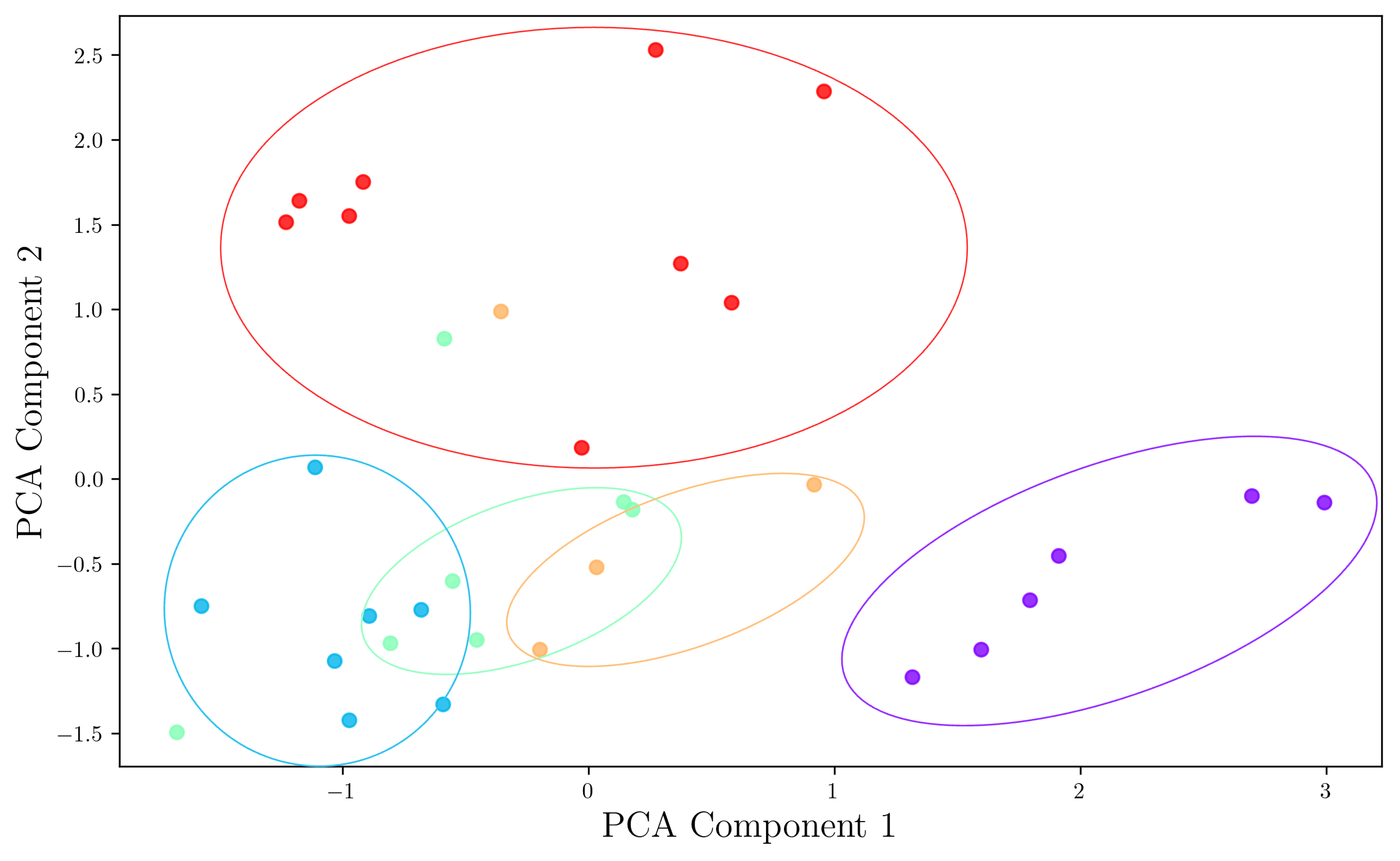}

\caption{Distribution of GQDs with oscillator strengths greater than 0.5 along the two principal components of the GQD data. The specific types of GQDs corresponding to each point are provided in the Supplementary Material.}

\label{pca}
\end{figure*}

In the principal component analysis (PCA) of GQD data (Figure \ref{pca}), we observe that the first two principal components (PC1 and PC2) together account for nearly half of the total variance in the dataset, with PC1 explaining 26\% and PC2 explaining 24\%. The complete list of PCA loadings are provided in the Supplementary Material.

PC1 is heavily influenced by oscillator strength, which has the highest loading (0.625). This suggests that oscillator strength is a primary factor in separating GQDs along PC1. Higher oscillator strengths correlate with GQDs positioned more positively along this axis, while those with lower oscillator strengths tend to cluster more negatively. Dopant percentage also plays a crucial role, with a loading of 0.555. GQDs with higher dopant percentages are distributed along PC1, contributing to the formation of distinct clusters. Interestingly, smaller GQD sizes (e.g., 1 nm) also contribute positively to PC1, suggesting that size plays a key role in distinguishing between clusters.

PC2, on the other hand, is primarily driven by emission energy, which has the highest loading (0.708), indicating that emission energy is a major factor differentiating GQDs along this axis. This reinforces the idea that GQDs with distinct emission properties, critical for optoelectronic applications, are clustered separately along PC2. Dopant percentage is also significant for PC2, but its negative loading (-0.453) suggests that lower dopant concentrations are associated with positive values along this axis. Structural features, particularly shape, are important in PC2 as well. Amorphous shapes have a high loading (0.328), suggesting that GQDs with more irregular structures cluster together, while square-shaped GQDs contribute significantly with a loading of 0.231, indicating a clear separation based on structural symmetry.

The PCA plot in Figure \ref{pca} illustrates these insights, with the five distinct clusters representing GQD samples separated by both optical and structural properties. For example, GQDs in the red cluster are likely characterized by higher oscillator strengths and emission energies, as they are positioned positively along PC1 and PC2. These GQDs could be prime candidates for applications requiring strong emission properties, such as in light-emitting devices. Meanwhile, the blue cluster at the lower-left of the plot likely represents GQDs with lower oscillator strengths and dopant concentrations, possibly making them better suited for applications where such characteristics are less critical. The separation of amorphous GQDs from more structured ones, such as the square and hexagonal shapes, suggests that shape plays a critical role in tuning the emission properties, likely through quantum confinement effects.

\section{Conclusions}

We systematically investigated the emission properties of 284 distinct graphene quantum dots (GQDs) using high-throughput TDDFT calculations and machine learning techniques, revealing key design rules for controlling emission through dopant type, concentration and GQD shape. Our findings indicate that sulfur doping consistently enhances emission energies, particularly at higher concentrations and facilitates emission within the visible spectrum, making it a promising candidate for optoelectronic applications. The analysis shows that quantum confinement effects, especially the size of the GQDs, play a crucial role in shifting the emission energy towards the infrared region as the structure size increases. Symmetry effects also contribute significantly, with hexagonal and square GQDs exhibiting distinct emission behavior due to their point group symmetries.

Hierarchical clustering demonstrated the influence of structural and chemical features, clustering GQDs based on shape, size and dopant characteristics. In particular, sulfur-doped amorphous GQDs formed distinct clusters, suggesting a unique interaction between sulfur doping and the amorphous structure, which could be leveraged in applications requiring customized emission properties. 

Principal component analysis (PCA) revealed that GQDs with high oscillator strength and higher emission energies clustered together, indicating their suitability for applications in high-efficiency light-emitting devices. The shape and size of the GQDs also influenced clustering, further reinforcing the importance of geometry in tuning optical properties. This study provides a comprehensive framework for the design of GQDs with tailored emission properties, offering practical guidelines for their development in optoelectronics, bioimaging and sensing applications.

\section{Acknowledgement}

\c{S}.\"{O}. is supported
by T\"{U}B\.{I}TAK under grant no. 120F354. Computing resources used in this
work were provided by the National Center for High Performance Computing
of T\"{u}rkiye (UHeM) under grant no. 1007872020 and TUBITAK ULAKBIM, High Performance and Grid Computing Center (TRUBA).

\bibliographystyle{naturemag}
\bibliography{grapheneEmission}


\end{document}


\title{Supplementary Materials: Emission Spectrum of Doped Graphene Quantum Dots: A high-throughput TDDFT Analysis}

\author{Mustafa Co\c{s}kun \"{O}zdemir}
\affiliation{Department of Chemistry, \.{I}zmir Institute of Technology,
\.{I}zmir, T\"{u}rkiye}

\author{Caner \"{U}nl\"{u}}
\affiliation{Department of Chemistry, \.{I}stanbul Technical University, \.{I}stanbul,
T\"{u}rkiye}

\author{\c{S}ener \"{O}z\"{o}nder}
\affiliation{Institute for Data Science \& Artificial Intelligence, Bo\u{g}azi\c{c}i University,
\.{I}stanbul, T\"{u}rkiye}
\email{Corresponding author: sener.ozonder@bogazici.edu.tr}

\maketitle

\begin{figure*}[h]
\includegraphics[scale=0.5]{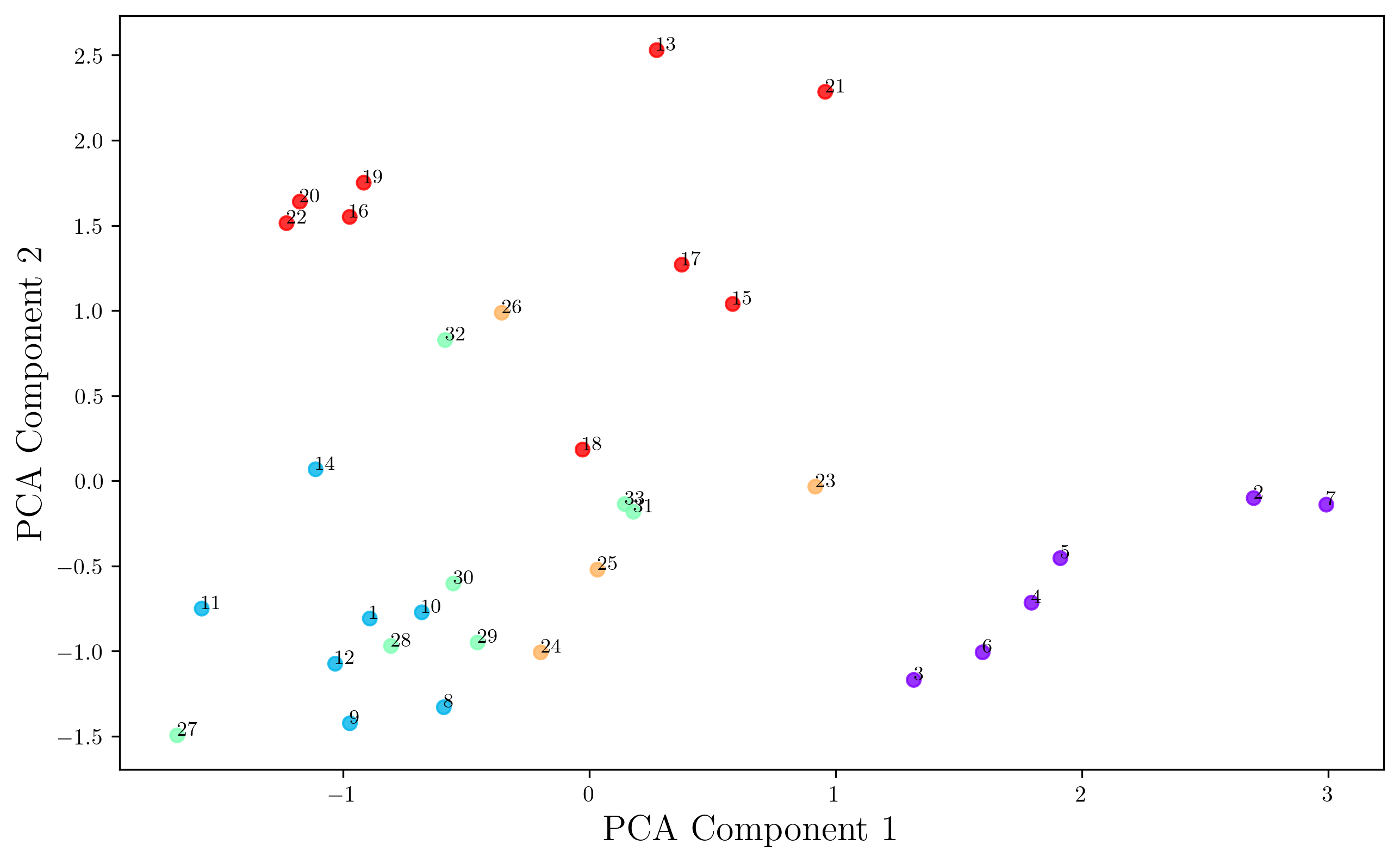}
\caption{The graphene quantum dot structures whose emission energies calculated via time-dependent density function theory.}
\label{pca-numbered}
\end{figure*}

The PCA plot given in Fig. \ref{pca-numbered} shows all 33 GQD structures whose oscillator strength is greater than 0.5. The corresponding structure identifiers are as follows: 1:1.5nm-Sqr-BN-3\%, 2:1nm-Sqr-B-5\%, 3:1nm-Sqr-BN-5\%, 4:1nm-Sqr-BP-5\%, 5:1nm-Sqr-NP-5\%, 6:1nm-Sqr-NS-5\%, 7:1nm-Sqr-P-5\%, 8:2nm-Sqr-B-3\%, 9:2nm-Sqr-N-3\%, 10:2nm-Sqr-NO-1.5\%, 11:2nm-Sqr-NP-1.5\%, 12:2nm-Sqr-NS-1.5\%, 13:Amo4, 14:Amo4-B-1.5\%, 15:Amo4-OS-3\%, 16:Amo4-S-1.5\%, 17:Amo4-S-3\%, 18:Amo4-S-5\%, 19:Amo1-B-1.5\%, 20:Amo1-N-1.5\%, 21:Amo1-P-1.5\%, 22:Amo1-SP-1.5\%, 23:Hex2-BN-5\%, 24:Hex2-BP-5\%, 25:Hex2-P-5\%, 26:Hex2-S-3\%, 27:Hex3-BP-3\%, 28:Hex3-NO-3\%, 29:Hex3-NS-3\%, 30:Hex3-O-3\%, 31:Hex3-OS-3\%, 32:Hex3-S-1.5\%, 33:Hex3-S-3\%

\subsection*{PCA Loadings for the First Two Principal Components}

The PCA loadings are given in Table \ref{pca-loadings}.
\begin{table}[h!]
\centering
\begin{tabular}{lcc}
\toprule
\textbf{Variable} & \textbf{PC1} & \textbf{PC2} \\
\midrule
Energy\_eV               &  0.407253 &  0.708410 \\
osc\_strength            &  0.624854 &  0.042717 \\
dopant\_percentage       &  0.554833 & -0.453267 \\
size\_1.5nm              & -0.019260 & -0.019194 \\
size\_1nm                &  0.266327 & -0.085160 \\
size\_2nm                & -0.104777 & -0.127014 \\
size\_amorph1            & -0.051046 &  0.170690 \\
size\_amorph4            & -0.018958 &  0.157638 \\
size\_hexagonal2         &  0.008660 & -0.013665 \\
size\_hexagonal3         & -0.080947 & -0.083295 \\
dopant\_B                &  0.001774 &  0.009214 \\
dopant\_BN               &  0.029099 & -0.047764 \\
dopant\_BP               & -0.001583 & -0.076388 \\
dopant\_N                & -0.046417 &  0.005133 \\
dopant\_NO               & -0.032090 & -0.041347 \\
dopant\_NP               &  0.007355 & -0.028595 \\
dopant\_NS               &  0.002447 & -0.071960 \\
dopant\_O                & -0.011925 & -0.014315 \\
dopant\_OS               &  0.016487 &  0.020353 \\
dopant\_P                &  0.086177 &  0.038551 \\
dopant\_S                & -0.030681 &  0.111123 \\
dopant\_SP               & -0.026571 &  0.035937 \\
dopant\_pristine         &  0.005929 &  0.060057 \\
shape\_new\_Amorphous    & -0.070003 &  0.328328 \\
shape\_new\_Hexagonal    & -0.072287 & -0.096960 \\
shape\_new\_Square       &  0.142290 & -0.231368 \\
\bottomrule
\end{tabular}
\caption{PCA Loadings for the First Two Principal Components.}
\label{pca-loadings}

\end{table}

\subsection*{GQD Structures and Emission Energies}

This section presents the 284 distinct graphene quantum dots (GQDs) of varying shapes (square, hexagonal, and amorphous) and sizes (~1-2 nm) that were analyzed using time-dependent density functional theory (TDDFT) to calculate their emission energies. These GQDs are doped with one or two elements from B, N, O, S, and P at dopant concentrations of 1.5\%, 3\%, 5\%, and 7\%. Among these, 12 specific structures exhibit emission energies within the visible spectrum, which are highlighted and presented in a dedicated table within the main article.

\begin{figure*}[!t]
\includegraphics[scale=0.45]{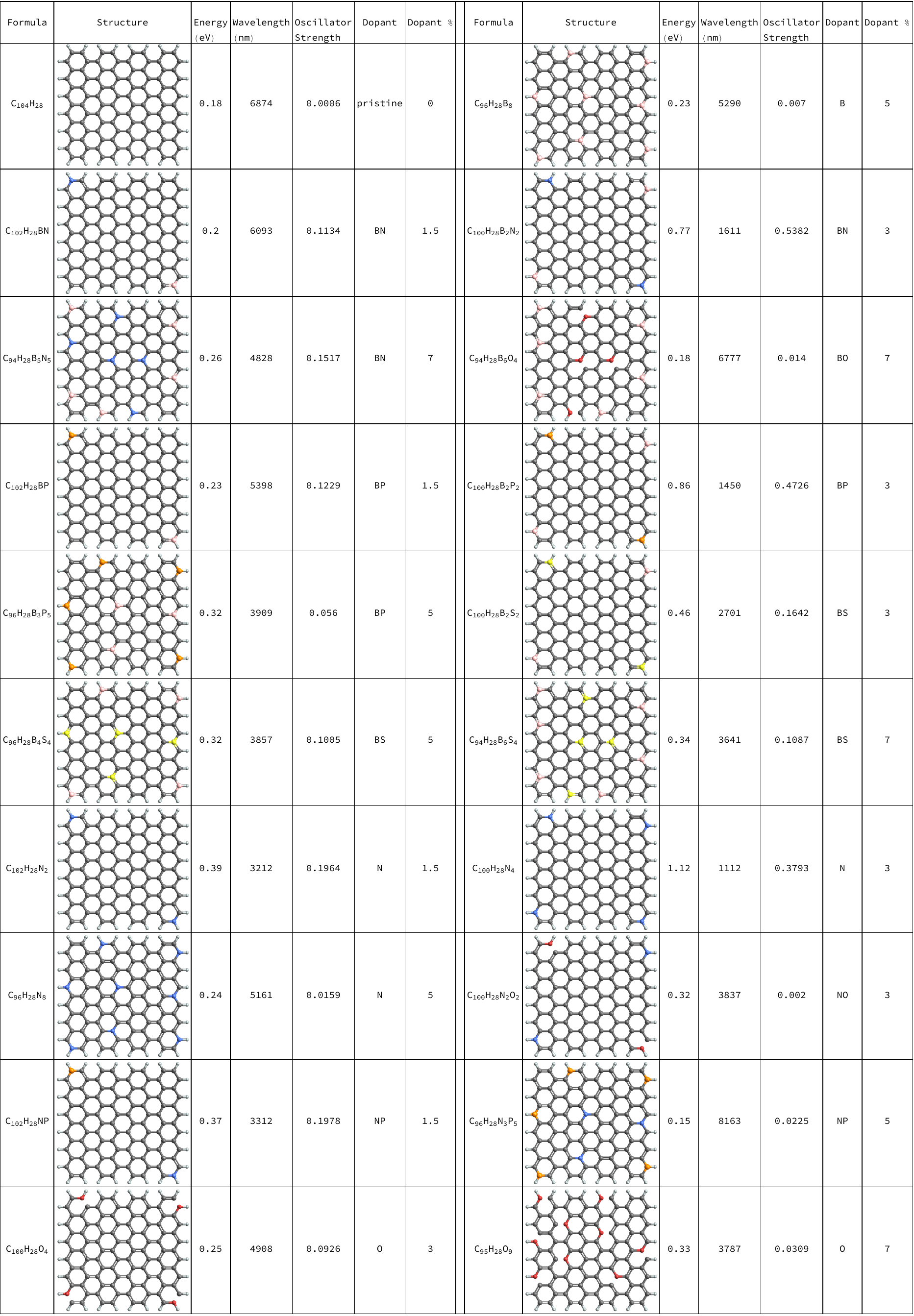}
\caption{The graphene quantum dot structures whose emission energies calculated via time-dependent density function theory.}
\end{figure*}

\begin{figure*}[!t]
\includegraphics[scale=0.45]{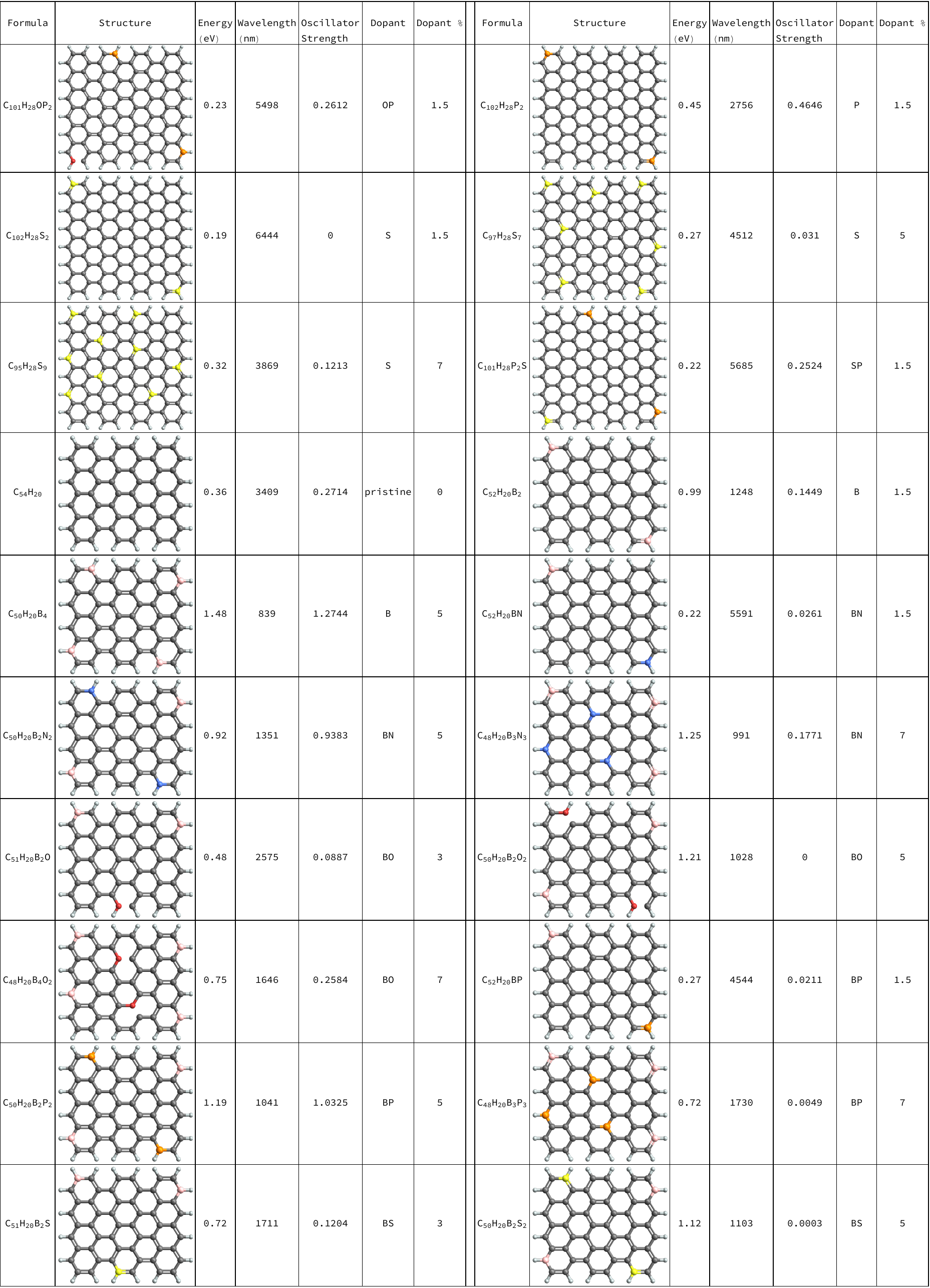}
\caption{The graphene quantum dot structures whose emission energies calculated via time-dependent density function theory.}
\end{figure*}

\begin{figure*}[!t]
\includegraphics[scale=0.45]{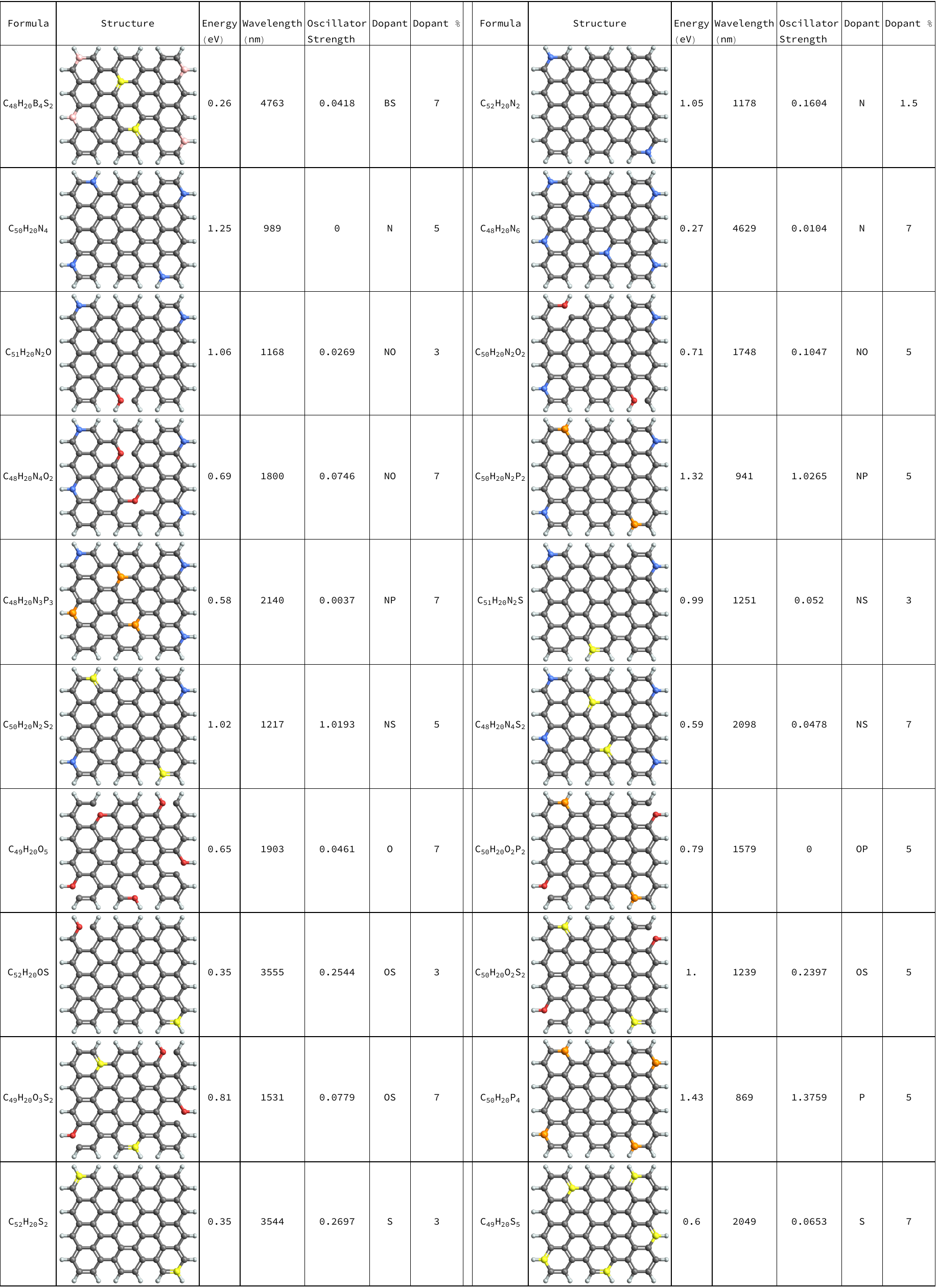}
\caption{The graphene quantum dot structures whose emission energies calculated via time-dependent density function theory.}
\end{figure*}

\begin{figure*}[!t]
\includegraphics[scale=0.45]{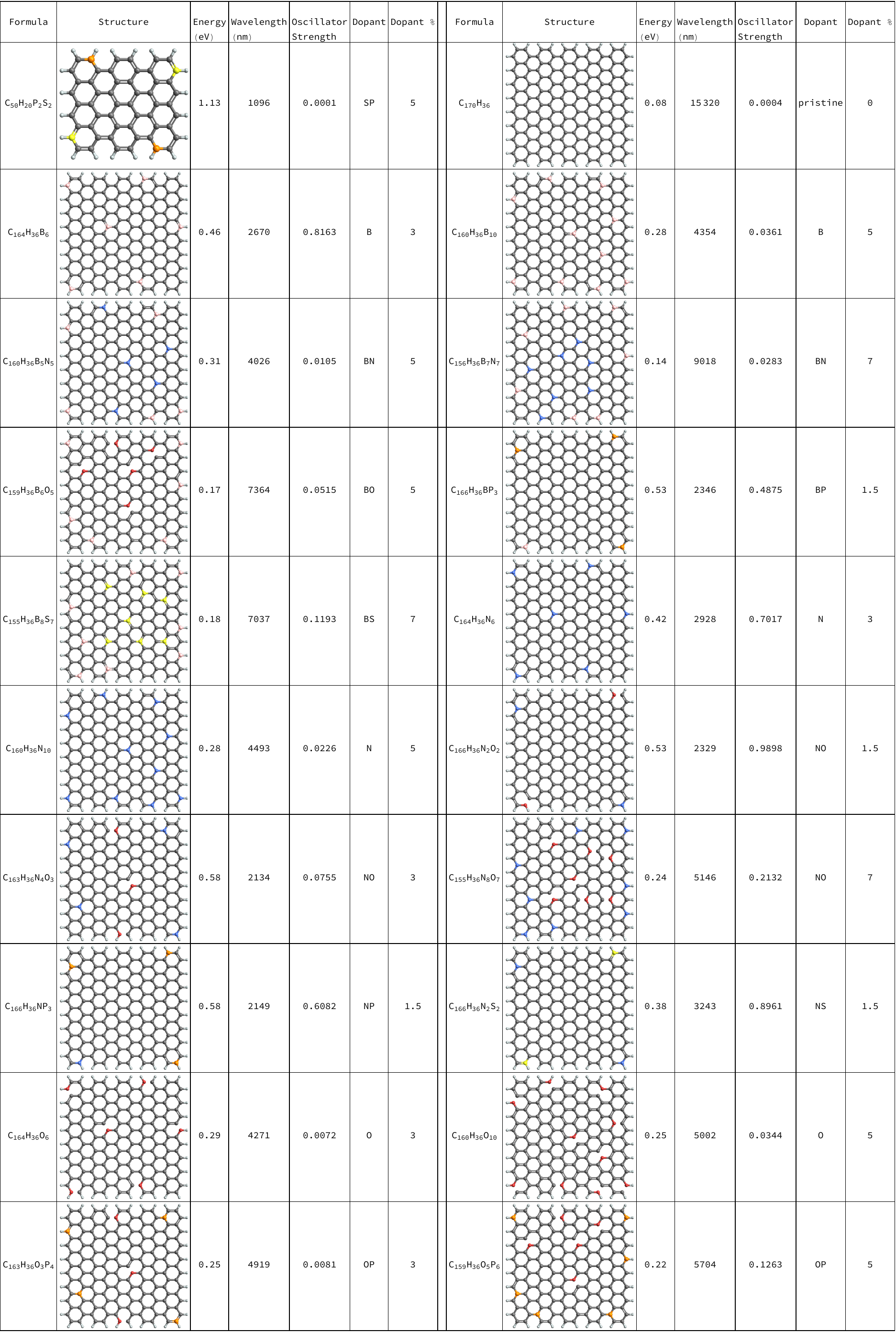}
\caption{The graphene quantum dot structures whose emission energies calculated via time-dependent density function theory.}
\end{figure*}

\begin{figure*}[!t]
\includegraphics[scale=0.4]{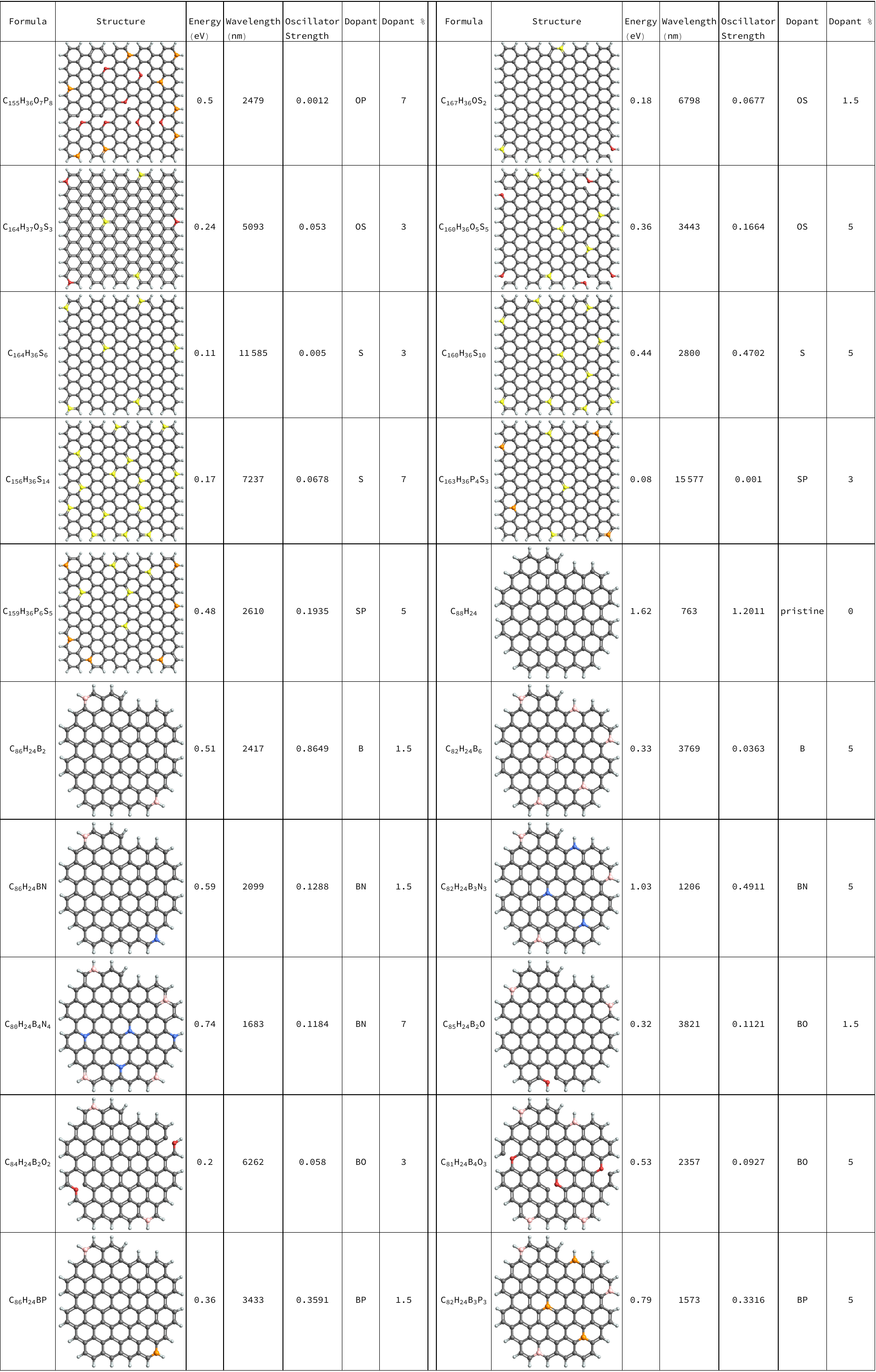}
\caption{The graphene quantum dot structures whose emission energies calculated via time-dependent density function theory.}
\end{figure*}

\begin{figure*}[!t]
\includegraphics[scale=0.4]{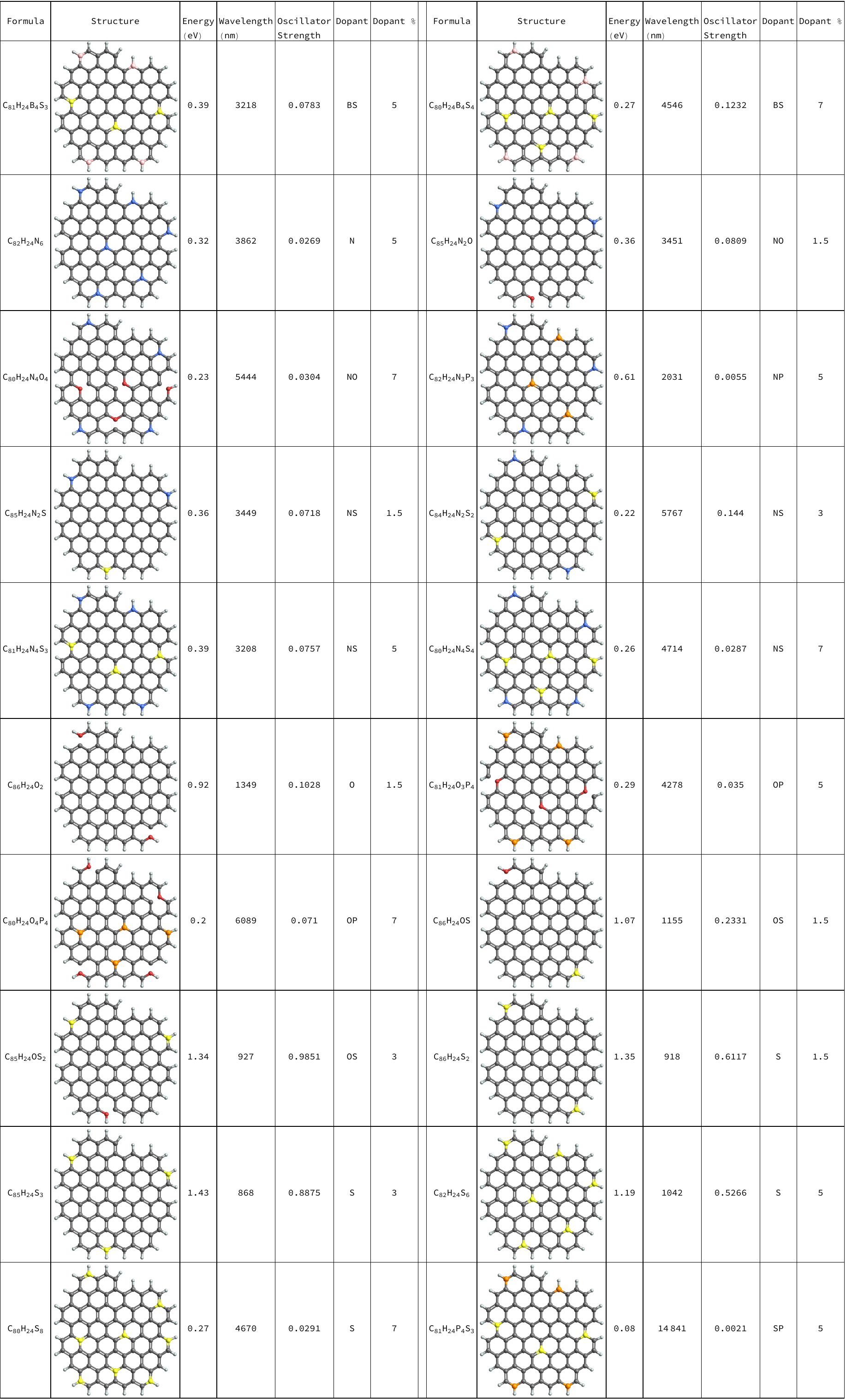}
\caption{The graphene quantum dot structures whose emission energies calculated via time-dependent density function theory.}
\end{figure*}

\begin{figure*}[!t]
\includegraphics[scale=0.45]{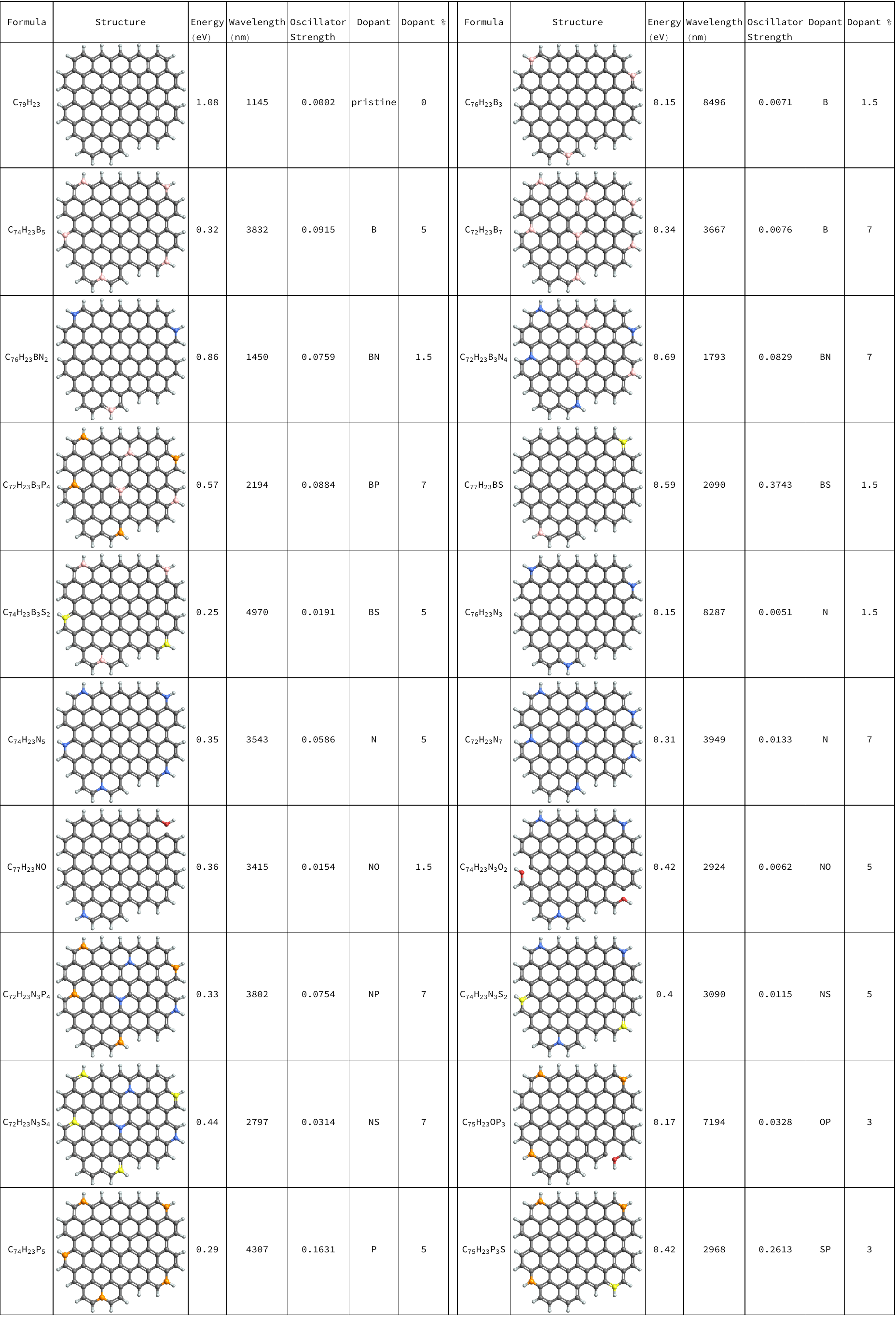}
\caption{The graphene quantum dot structures whose emission energies calculated via time-dependent density function theory.}
\end{figure*}

\begin{figure*}[!t]
\includegraphics[scale=0.42]{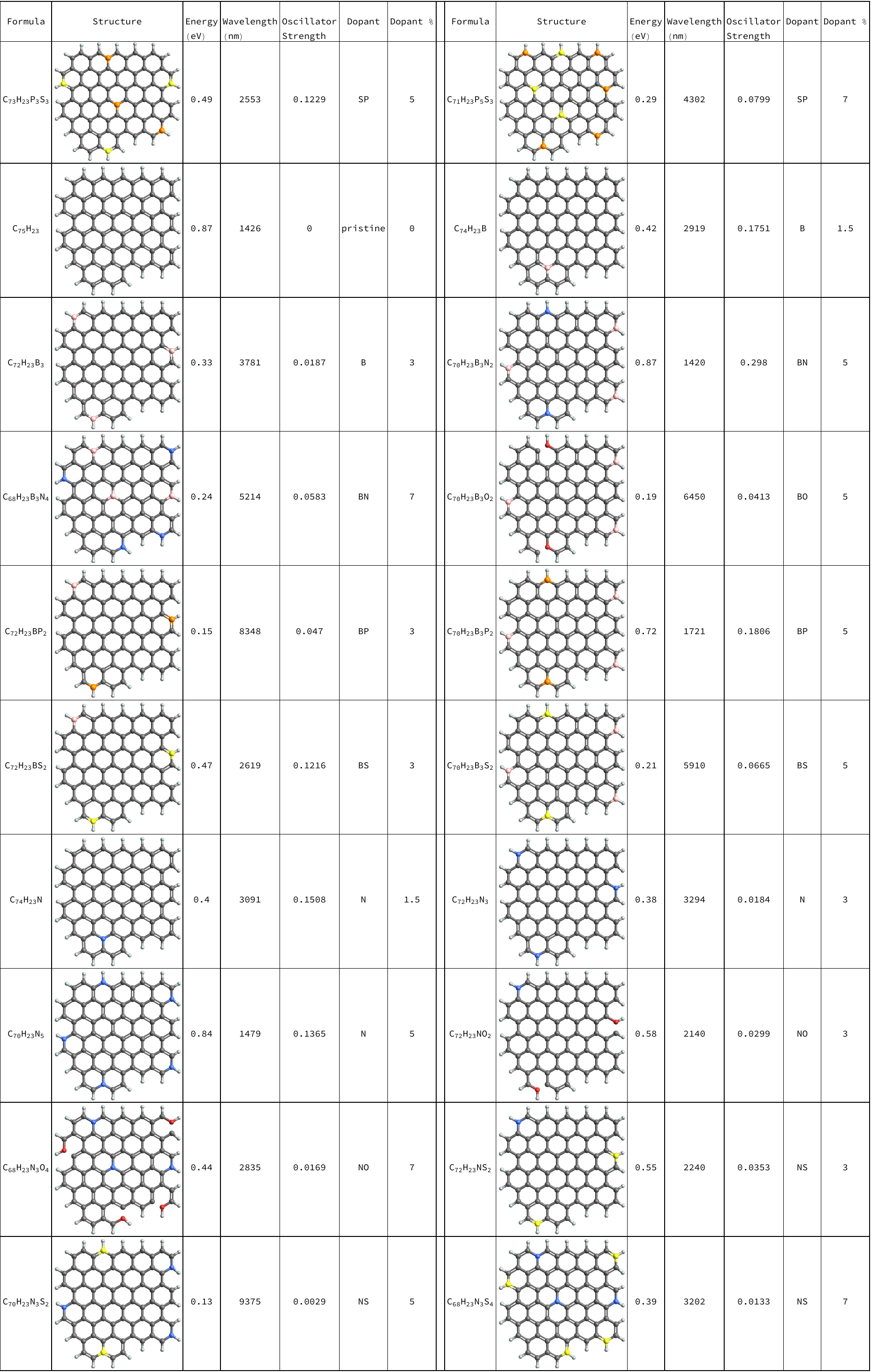}
\caption{The graphene quantum dot structures whose emission energies calculated via time-dependent density function theory.}
\end{figure*}

\begin{figure*}[!t]
\includegraphics[scale=0.35]{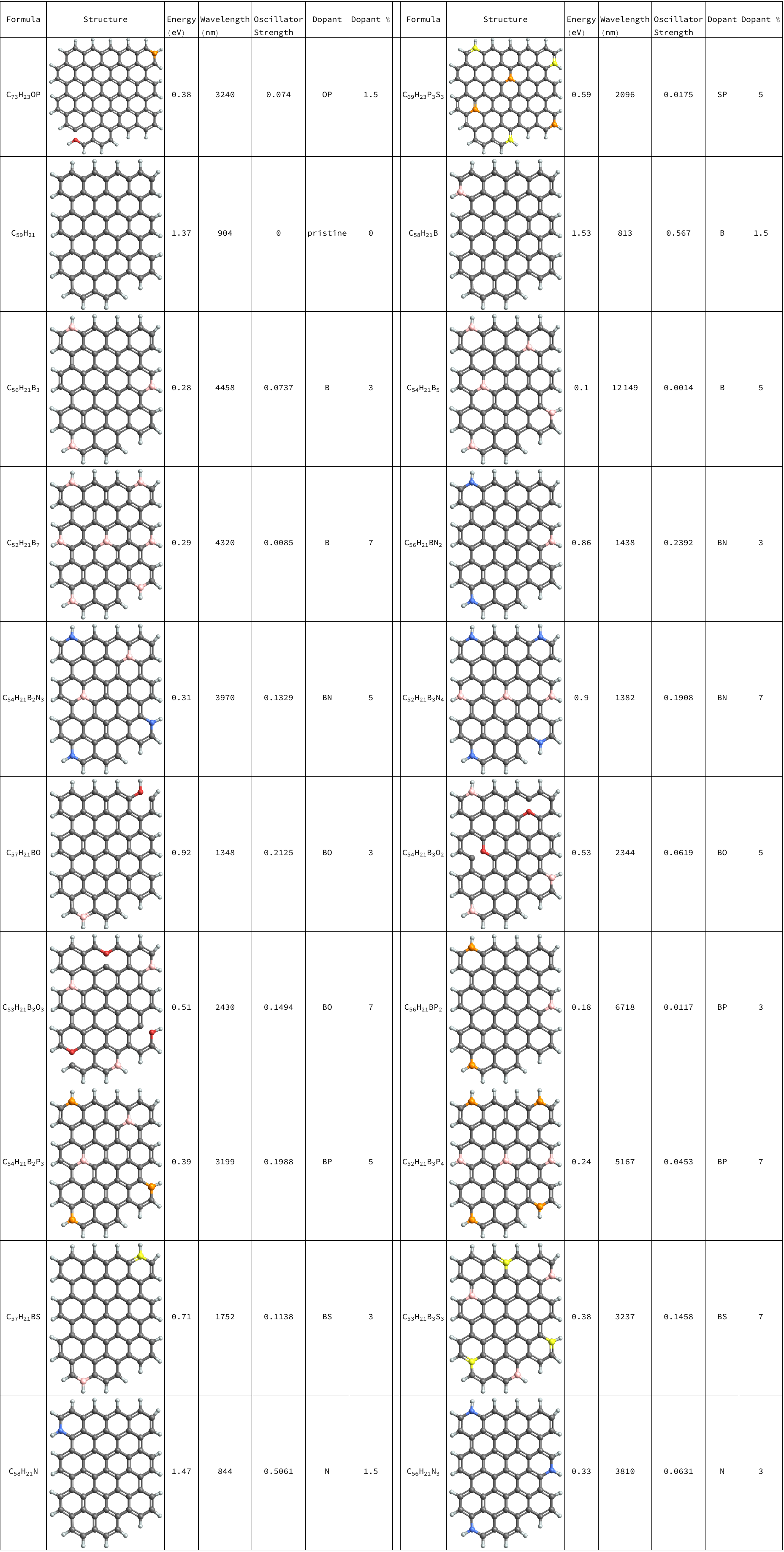}
\caption{The graphene quantum dot structures whose emission energies calculated via time-dependent density function theory.}
\end{figure*}

\begin{figure*}[!t]
\includegraphics[scale=0.37]{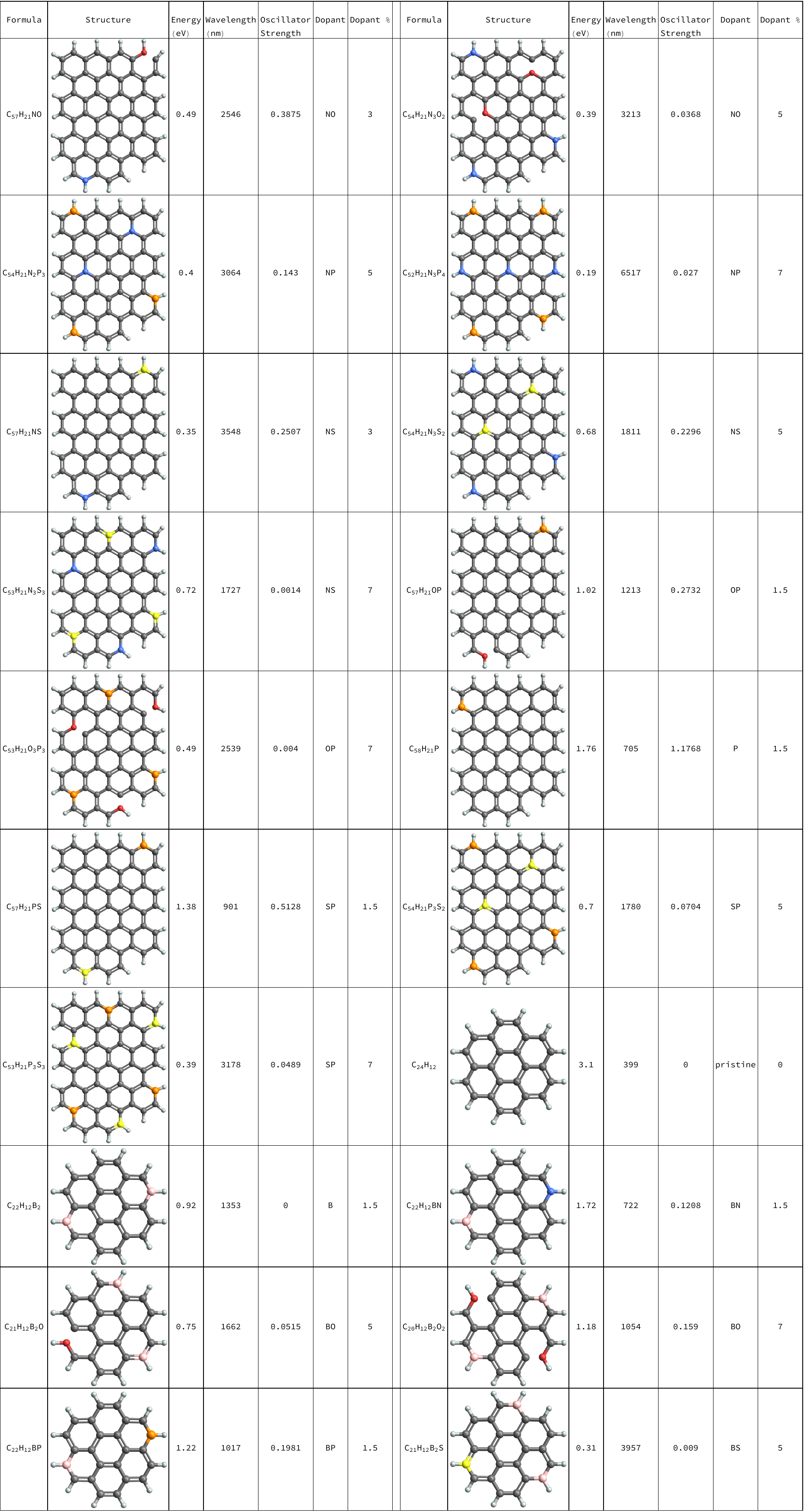}
\caption{The graphene quantum dot structures whose emission energies calculated via time-dependent density function theory.}
\end{figure*}

\begin{figure*}[!t]
\includegraphics[scale=0.45]{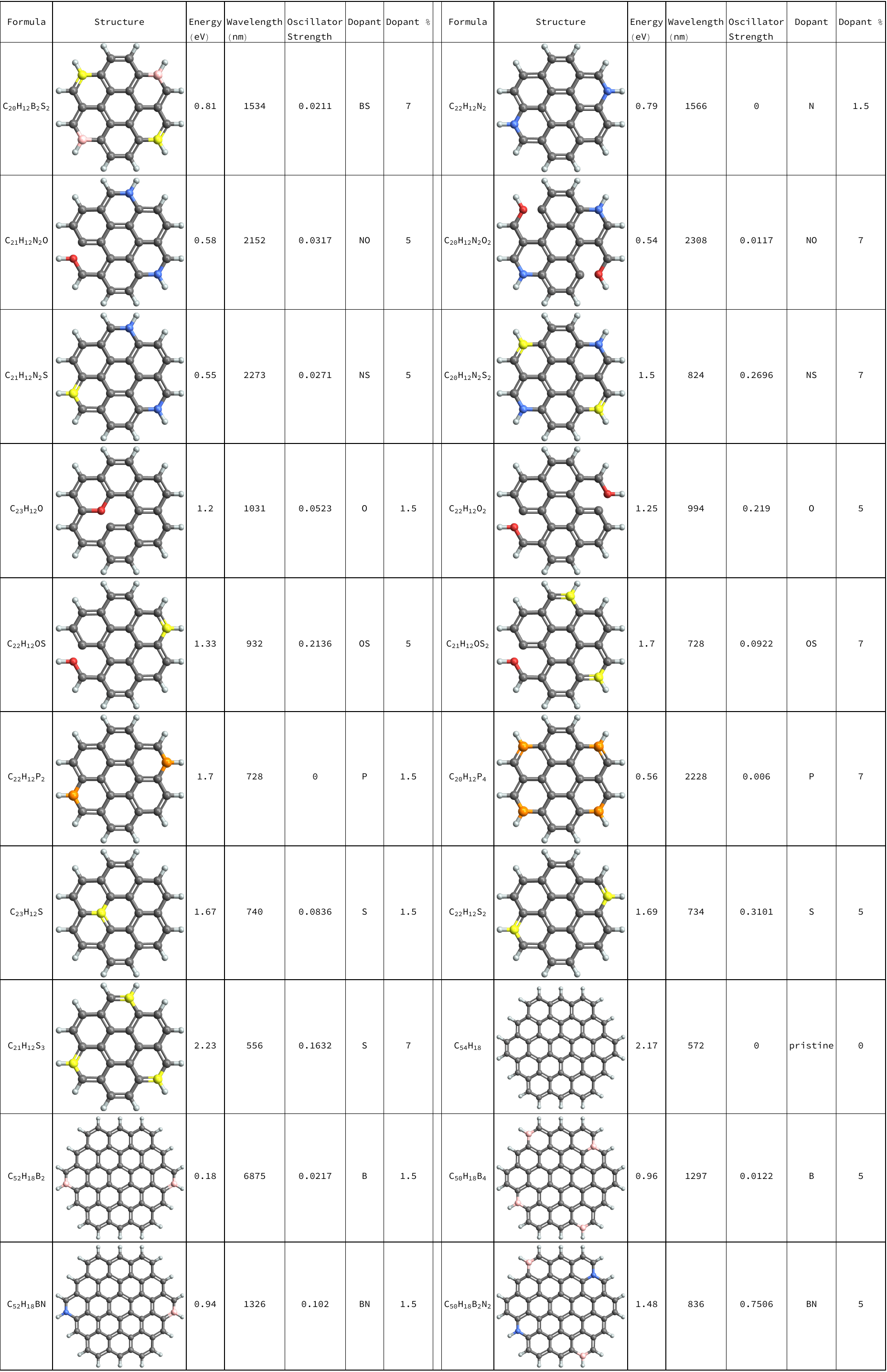}
\caption{The graphene quantum dot structures whose emission energies calculated via time-dependent density function theory.}
\end{figure*}

\begin{figure*}[!t]
\includegraphics[scale=0.45]{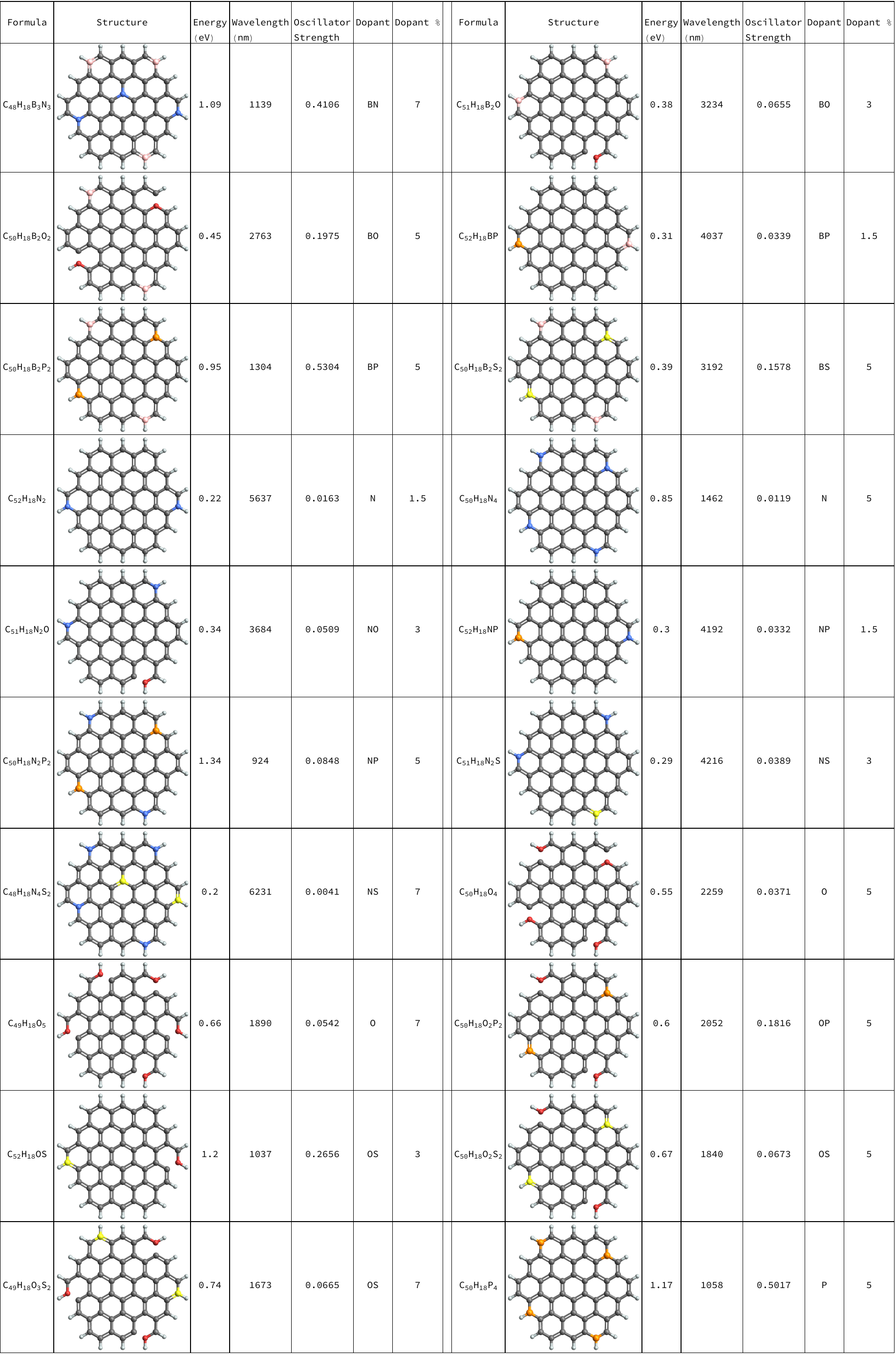}
\caption{The graphene quantum dot structures whose emission energies calculated via time-dependent density function theory.}
\end{figure*}

\begin{figure*}[!t]
\includegraphics[scale=0.45]{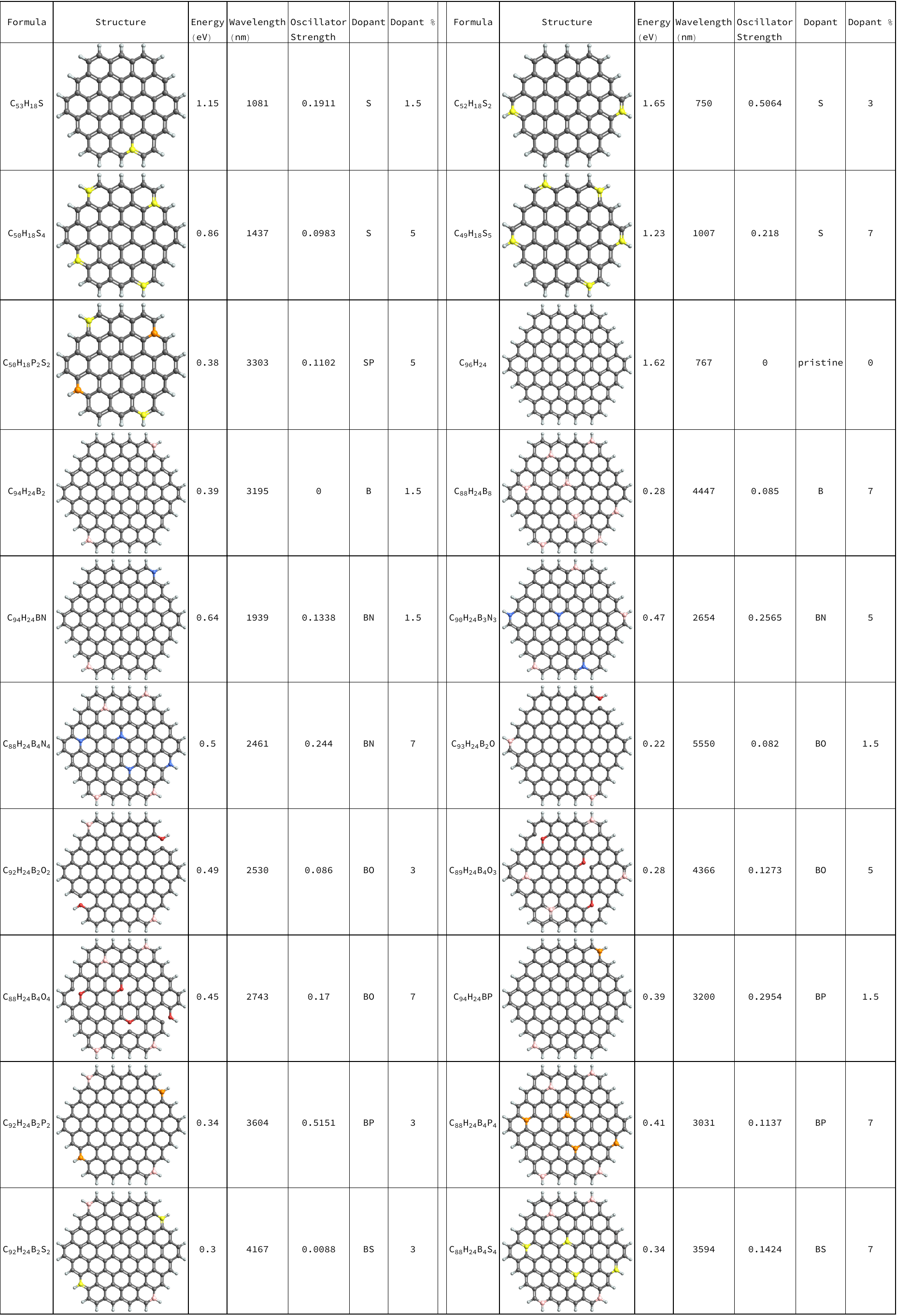}
\caption{The graphene quantum dot structures whose emission energies calculated via time-dependent density function theory.}
\end{figure*}

\begin{figure*}[!t]
\includegraphics[scale=0.45]{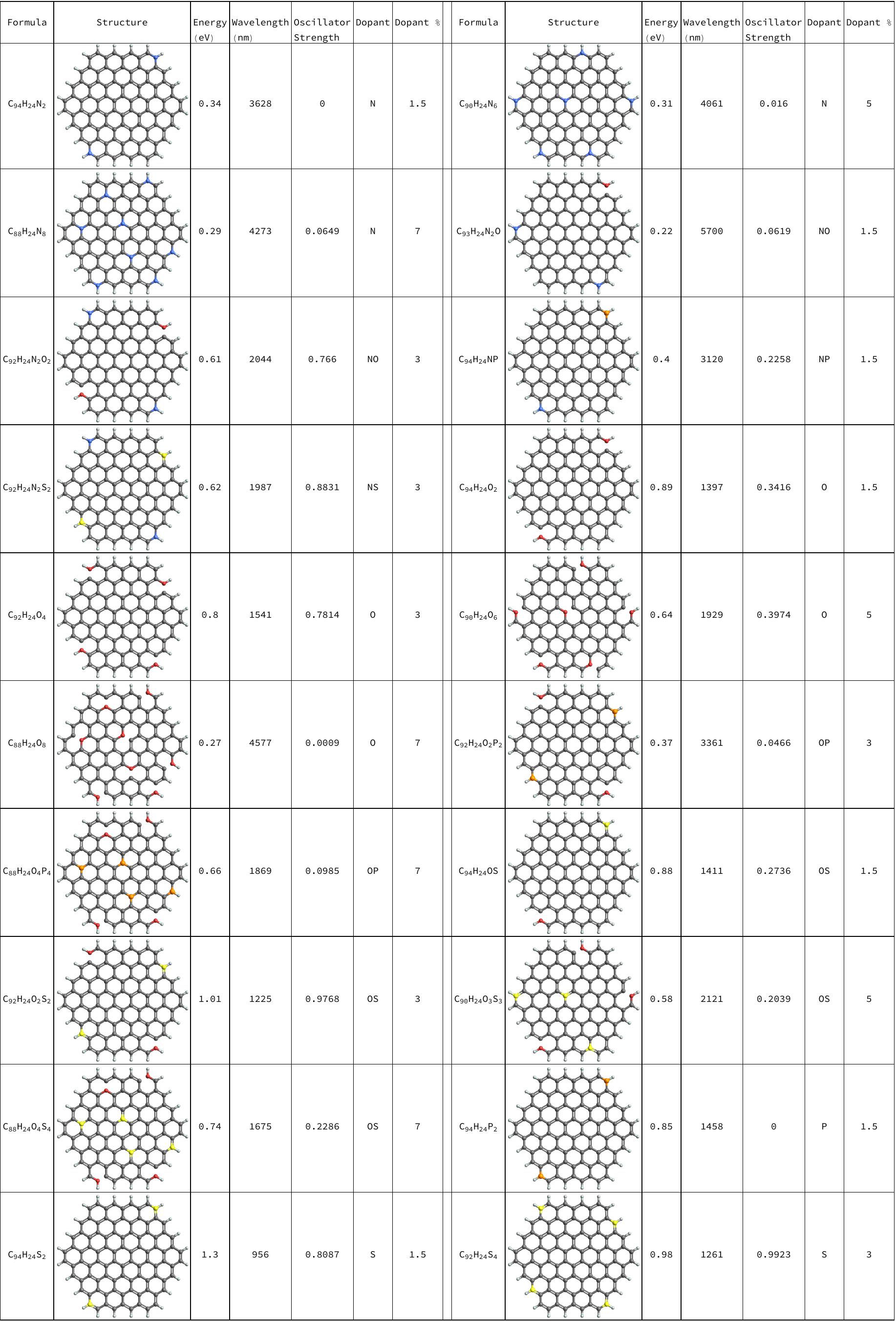}
\caption{The graphene quantum dot structures whose emission energies calculated via time-dependent density function theory.}
\end{figure*}

\begin{figure*}[!t]
\includegraphics[scale=0.45]{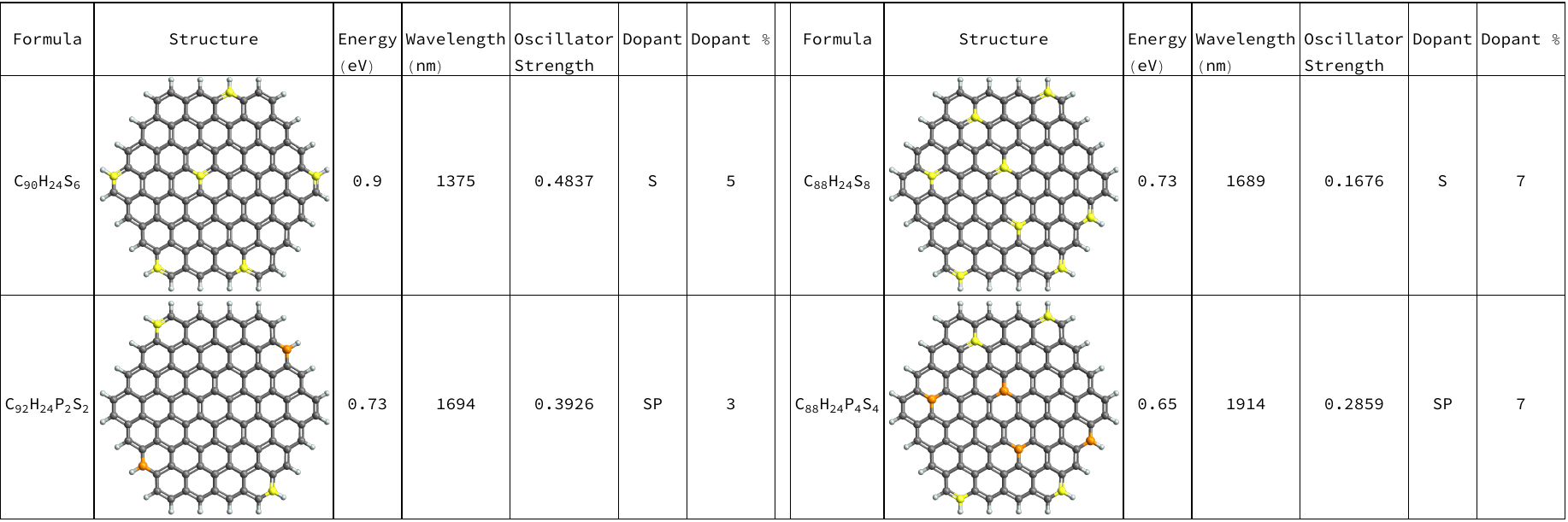}
\caption{The graphene quantum dot structures whose emission energies calculated via time-dependent density function theory.}
\end{figure*}